\documentclass[usenatbib,usegraphicx]{mn2e}
\usepackage{textcomp}
\usepackage{amsmath}
\include{bib_names}
\title[Optical polarisation of the Crab pulsar]{Optical polarisation
of the Crab pulsar: precision measurements and comparison to the radio
emission}
\author[A. S\l{}owikowska et al.]{A. S\l{}owikowska$^{1,2}$\thanks{E-mail:
aga@physics.uoc.gr (FORTH); aga@ncac.torun.pl (NCAC)},
G. Kanbach$^{3}$, M. Kramer$^{4}$ and A. Stefanescu$^{3}$\\
$^{1}$IESL, Foundation for Research and Technology - Hellas, P.O. Box 1385,
71110 Heraklion, Crete, Greece\\
$^{2}$Nicolaus Copernicus Astronomical Center, Rabia\'nska 8, 87-100 Toru\'n,
Poland\\
$^{3}$Max-Planck Institut f\"ur Extraterrestrische Physik, 85741 Garching bei
M\"unchen, Germany\\
$^{4}$University of Manchester, Jodrell Bank Observatory, Macclesfield, Cheshire
SK11 9DL, UK}
\begin{document}

\date{Accepted . Received ... ; in original form }

\pagerange{\pageref{firstpage}--\pageref{lastpage}} \pubyear{2008}

\maketitle

\label{firstpage}

\begin{abstract}
The linear polarisation of the Crab pulsar and its close
environment was derived from observations with the high-speed
photo-polarimeter OPTIMA at the 2.56-m Nordic Optical Telescope in
the optical spectral range (400--750~nm). Time resolution
as short as 11 $\mu$s, which corresponds to a phase interval
of 1/3000 of the pulsar rotation, and high statistics allow
the derivation of polarisation details never achieved before.
The degree of optical polarisation and the position angle
correlate in surprising details with the light curves at optical
wavelengths and at radio frequencies of 610 and 1400 MHz. Our
observations show that there exists a subtle connection between
presumed non-coherent (optical) and coherent (radio) emissions.
This finding supports previously detected correlations between
the optical intensity of the Crab and  the  occurrence  of  giant
radio  pulses. Interpretation  of  our observations require
more elaborate  theoretical models than those currently available
in the literature. 
\end{abstract}


\section{Introduction}

\begin{table*}
\caption{High energy polarisation measurements of pulsars and
their nebulae. Polarisation degree is given as a percentage, and
position angle in degrees (North, $0^\circ$ to East, $90^\circ$).}
\label{Tab:pol_summary}
\begin{tabular}{@{} l c c r}
\hline
\hline
Band & Pulsar / Polarisation & Nebula (near PSR) & Ref.\\

\hline
 & \textbf{Crab} & & \\

Optical (V$\sim 16.6$)  & phase-resolved  (Fig.~\ref{Fig:pa_pd}) &
$9.7\%\pm0.1\%$ ; $139.8\degr\pm0.2\degr$  & [1], [2], [3] \\
 & phase-averaged $9.8\%\pm0.1\%$  and $109.5\degr\pm0.2\degr$
 &  ($< 5\arcsec$ from PSR) &  \\
 & & & \\

UV  & phased-resolved, similar to optical -
 & & [4] \\
 & P.D. and P.A. (MP and IP)  &  & \\
 & & & \\

X-ray  & only upper limits &
$19.2\%\pm0.9\%$; $155.8\degr\pm1.4\degr$   & [5], [6] \\
 & & & \\

Hard X-ray/ soft $\gamma$-ray & off-pulse (phase: 0.52--0.88) $>72\%$; 
$120.6\degr \pm 8.5\degr$ &   &   [7] \\
 & phase-averaged $47\%^{+19\%}_{-13\%}$; $100\degr \pm 11\degr$ & & [7] \\
 & & & \\

$\gamma$-ray & off-pulse (phase: 0.5--0.8) $46\%\pm10\%$;
$123\degr\pm11\degr$  & & [8] \\
\hline
 &  \textbf{B0540-69} & & \\

Optical (V$\sim 22.5$) & phase-averaged: $\sim 5\%$, no error
quoted; phase-resolved: $< 15\%$ &
$5.6\% \pm 1.0\%$ ; $79\degr \pm 5\degr$ & [9], [10], [11] \\

\hline
 & \textbf{Vela} & & \\

Optical (V$\sim 23.6$) & phase averaged: $9.4\% \pm 4\% $,
$146\degr \pm 11\degr$ & & [9], [12] \\

\hline
 & \textbf{B0656+14} & & \\

Optical (V$\sim 25$)  & double peak light curve;
P.D. bridge $\sim$ 100\%, peaks $\sim$ 0\% & & [13] \\
 &  P.A. sweeps in agreement with RVM & & \\

 \hline
 & \textbf{B1509-58} & & \\

Optical (V$\sim 25.7$) & phase-averaged: $\sim 10.4\%$ (very
uncertain, no error quoted) & & [9]\\

\hline
\end{tabular}

\medskip
[1] \citet{Smith1988}, [2] \citet{Slowikowska2008}, [3] this paper,
[4] \citet{Graham-Smith1996},
[5] \citet{Silver1978}, [6] \citet{Weisskopf1978}, [7] \citet{Forot2008},
[8] \citet{Dean2008}, [9] \citet{Wagner2000}, [10] \citet{Middleditch1987},
[11] \citet{Chanan1990}, [12] \citet{Mignani2007}, [13] \citet{Kern2003}, 
\end{table*}

The Crab pulsar and its pulsar wind nebula (PWN) are two of the
most intensively studied objects in the sky. The compact remnant
of SN1054, a cornerstone of high energy astrophysics, is one of
the youngest and most energetic pulsar and its pulsed emission
has now been detected throughout the electromagnetic spectrum from
about 10 MHz \citep{Bridle1970} up to $>25$ GeV with some evidence
($3.4\sigma$) of pulsed emission above 60~GeV \citep{Aliu2008}.
The PWN was measured even up to energies of $\sim 100$ TeV
\citep{Aharonian2004, Allen2007PhD, Allen2007}. The pulsar and
its nebula are predominantly sources of non-thermal radiation
(synchrotron, curvature and inverse Compton processes), which is
indicated not only by the broad-band spectral continua but also by
the strong polarisation of these emissions. The outstanding
brightness of the Crab across the electromagnetic spectrum makes it
the ideal target to investigate polarisation at all wavelengths
where suitable instrumentation is available.

Before we focus on the Crab we want to introduce the subject of polarisation
in pulsars by a short review of the total population: most pulsars
were discovered in the radio regime (presently about two thousand objects
are listed in the Australia Telescope National Facility Pulsar
Catalogue\footnote{http://www.atnf.csiro.au/research/pulsar/psrcat/},
\citealt{Manchester2005}) and for  for a large fraction of them
polarisation studies were performed \citep[e.g.][]{Gould1998,Karastergiou2006}.
Most radio pulsars were found to show strong linear polarisation, including
often a characteristic a swing of the position angle (P.A.) in an S-like
shape near the pulse centre. This swing is interpreted in the `rotating vector
model' \citep[hereafter: RVM,][]{Radhakrishnan1969} as a projection of the
magnetic field line at the point of emission onto a plane perpendicular to the
observer's sight-line. The point of emission is usually assumed to be
in the polar cap region of the pulsar where a regular dipolar field
line points with a small angle (beam width) towards the observer.
The free parameters of this simple geometrical model are the
inclination angle between the axes of rotation and magnetic
dipole, and the viewing angle between the line of sight and the
rotation axis. Analyses of radio polarisation from many pulsars
\citep[e.g. references in][]{Lyne2006} showed that most sources can
be described in a polar cap model with an RVM and the intrinsic
polarisation could be as high as $100\% $. Several pulsars
however, especially at radio frequencies above several GHz, show
reduced or complex polarisation signatures with indications of
depolarisation or switching between modes of polarisation. These
complications could arise from the subtle and often highly
variable and unstable nature of the coherent processes underlying
the radio emission.

The situation changes when going to higher energies, i.e. optical,
X- and $\gamma$-ray regimes. Incoherent single particle radiation processes
in the magnetosphere provide the high-energy pulsar emission (thermal emission
from the neutron star surface is not considered here). Precise polarisation
measurements of polarisation degree (P.D.)\footnote{The terms: polarisation degree,
P.D. and $p$ as well as position angle, P.A. and $\theta$ are used
interchangeably.} and P.A. as a function of pulsar phase in a wide range of
energies should provide deep insight into the pulsars emission
mechanisms, particle spectra, and emission site topologies. Fast
X- and $\gamma$-ray polarimetry from space borne instruments is
presently of very limited sensitivity. Therefore results have  been reported
only for the brightest pulsar and its PWN, i.e. the Crab. In the optical domain the
situation is somewhat reversed: here the polarimeters are quite sensitive but
the optical magnitudes of most pulsars are very faint. Therefore, despite
the increasing number of optical pulsars (fourteen are known), only for five of
them attempts to measure the pulsar and/or nebular polarisation were made. Table
\ref{Tab:pol_summary} summarises these efforts. For three pulsars phase-resolved
measurements were performed; for the remaining two only phase-averaged results
are available. Again the Crab, being the brightest object, is the exception
with a multitude of optical polarisation measurements.

The scarcity of optical polarisation measurements, either in a
phase averaged mode or with time resolution of the pulsar
rotation, is clearly due to the large difference in magnitude
going from the brightest pulsar, i.e. the Crab, to the next,
B0540-69, which is $\sim 6$ magnitudes fainter. Therefore only for
the Crab pulsar fully phase resolved polarisation measurements
have been possible so far. The first optical phase-resolved linear
polarisation observations of the  Crab pulsar
\citep{Wampler1969,Cocke1970,Kristian1970}  showed that  the
polarisation angle  sweeps  through  each  peak and the
polarisation degree decreases  and then  increases within each
pulse, reaching the minimum shortly after the pulse peak. These
early observations were limited to the main pulse (MP) and inter
pulse (IP) phase ranges only. Interpretation of the polarisation
pattern behaviour has been made  in terms  of geometrical  models
\citep{Wampler1969, Radhakrishnan1969, Cocke1973, Ferguson1973, Ferguson1974},
some  of which offer the possibility of   determining  the   magnetospheric
location   of  the   radiation  source \citep[e.g.][]{Cocke1973,Ferguson1974}.

For a long time  it  was thought  that  optical radiation  of the
Crab pulsar persisted only through both peaks, the MP and IP, and in the bridge
region between them, but not in the phase range following the IP and preceding
the MP. Several phase resolved imaging observations \citep{Peterson1978,
Percival1993, Golden20002d}  showed however that  radiation persists throughout
the  whole  pulsar rotation.  They  found  that  the minimum  intensity,
consistent with  coming from an  unresolved source at the  pulsar position,
occurs  immediately  before the  start  of  the MP  within  a  phase range  of
approximately 0.779 and 0.845.
This level, often called the DC\footnote{Direct Current or continuous current,
in this case refers to a continuous emission component that could be present
throughout the whole rotational phase.} level of the Crab pulsar, has been
variously determined with an intensity relative to the MP maximum of $3.6\%$
\citep{Peterson1978}, $<0.9\% $ \citep{Percival1993}, or $\sim 1\% $
\citep{Golden2000off}.  

After this discovery, the linear polarisation of optical
radiation  from the Crab pulsar was measured by \citet{Jones1981} and
\citet{Smith1988}. Both results confirmed  the previous  observations and
additionally  showed the polarisation characteristics during both  off-pulse
phase ranges (bridge and DC). However,  for the DC phase range the results
were not conclusive. The polarisation  degree was  at a level of 70\% and
$47\pm10\%$  and the position  angle  was $\sim 118\degr$ and  $\sim 130\degr$ 
for \citet{Jones1981} and \citet{Smith1988}, respectively.

\citet{Weisskopf1978} measured the linear polarisation of the
X-ray flux from the Crab nebula at energies of 2.6 keV and 5.2
keV, independent of any contribution from the pulsar. Within a field
of view of $3\degr$ the X-rays are polarised at a level of 
$19.2\%-19.5\%$ and a P.A. of $156\degr-152\degr$. For the
pulsar itself only upper limits could be derived
\citep{Silver1978}. No evidence for an energy dependence of the X-ray
polarisation was found. Close to the pulsar the nebular optical
polarisation is  quite uniform (at $\sim 9\% - 11\%$) but the
position angles change steadily with radial distance; $2\arcsec - 3\arcsec$
from the pulsar the mean value is around $140\degr$ but beyond $5\arcsec$ the
position angle exceeds $155\degr$ and it is very
location-dependent \citep{McLean1983}. The X-ray results for the
nebula are in good agreement with the optical polarisation
measurements which yield a $19\%$ polarisation at $162\degr$ for
the central region of the nebula within $0\farcm486$ radius
\citep{Oort1956}. The similarity of the optical and X-ray results
indicates that the polarisation is independent of energy over a
broad  spectral range. Recently $\gamma$-ray phase-averaged
(off-pulse phase region, i.e. 0.5-0.8) polarisation measurements of
the Crab nebula and pulsar were reported by \citet{Dean2008} for the
energy range $100~\rm{kV} - 1~\rm{MeV}$. The P.D. was  found
at $46\%\pm10\%$ and the P.A. is $123\degr\pm11\degr$, which indicates that this
energetic radiation is dominated by emission from the inner nebula around the
pulsar. This results has been recently confirmed by \citet{Forot2008} by using
the IBIS instrument on board of the \textit{INTEGRAL} telescope to obtain
the linearly polarised emission for energies between $200 - 800~\rm{keV}$
and extend polarisation characteristics for pulses and bridge phases
(Tab.~\ref{Tab:pol_summary}). 

We structure the paper as follows: firstly we describe our
instrumentation (\S 2). In \S 3 and \S 4 the observations,
data reduction methodology and measurements of polarimetric
and photometric standards are described. In \S 5 we present
the results of the Crab nebula and its pulsar. Also in \S 5
we show the pulsar polarisation characteristics after
subtraction of an unpulsed component and discuss time alignment
between optical and radio wavelengths. In \S 6 we present a
discussion of theoretical models and compare them to our
measurements and end with a summary and conclusions.

\section{Instrumentation}
Our goal was to obtain polarisation characteristics
of the Crab pulsar with very high time resolution as a function of
the pulsar rotational phase. We used the 2.56-m Nordic Optical
Telescope\footnote{\texttt{http://www.not.iac.es/telescope/technical-details.html}}
(NOT) for these measurements. This facility not only  provides a  mirror
with reasonably large collecting area but  also has  an excellent
performance in offset  guidance and  telescope control, which is
important for instruments   with   very    small   entrance
apertures. Moreover, contrary  to  many  larger  optical
telescopes, guest instruments can be used at NOT.  For the Crab
pulsar  observations  carried out during November, 23--27 2003, we
used the high-speed photo-polarimeter
OPTIMA\footnote{\texttt{http://www.mpe.mpg.de/gamma/instruments/optima/}}
(Optical Pulsar TIMing Analyser). We shortly
describe the system below. An extended description of the OPTIMA
instrument is presented by \cite{Kanbach2003, Kanbach2008}.

\subsection{OPTIMA}
\label{Sec:OPTIMA}

\begin{figure}
\includegraphics[scale=0.33]{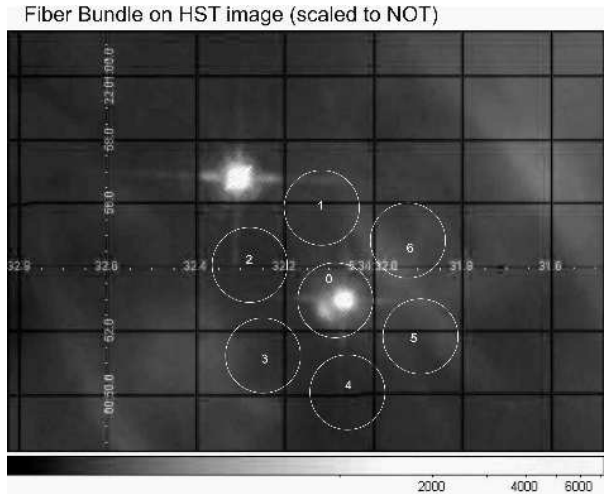}
\caption{An  enlargement  of  the  HST image  of  the inner Crab
nebula \citetext{source R. Romani, priv. comm.}. North is up; East is left. The  pulsar is
identified with the lower/right of the two stars near the
geometric centre of the nebula. A small
arc-like  feature (the inner knot) which is clearly resolved in
the HST image is located $0\farcs65$ to the SE of the
pulsar. The OPTIMA fibre bundle,
centred on the pulsar  and scaled with the NOT focal plane scale,
is over plotted.}
\label{Fig:HST}
\end{figure}

The basic requirement for a high-speed photometer is to select and time tag
individual photons from a celestial source with high efficiency and precision.
In the case of OPTIMA, which was primarily designed for studying the optical
light curves of faint pulsars and other highly variable targets, the source flux
is isolated  in the  focal plane  of the telescope by the use of an optical
fibre pick-up which acts as a diaphragm. The target fibre is central to a
hexagonal bundle of identical fibres which measure the sky background and,
in the case of the Crab nebula, the close environment of the pulsar
(Fig.~\ref{Fig:HST}). OPTIMA uses commercial single photon counting modules
with Avalanche Photodiodes (APDs)\footnote{\texttt{Perkin-Elmer SPCM-AQR-15FC}}.
These APD counters have quantum efficiencies peaking at 70\% for
$\lambda\sim 700$~nm and a wide range with Q.E. $>20\%$ from 440 to 980~nm.

The  light from the telescope at the Cassegrain focus is incident on a slant
mirror with an embedded bundle of optical fibres. Optionally one can insert
filters  or a rotating linear polariser into  the incoming beam. The field
around the fibres,  visible in the mirror (typical size $2\arcmin
\times 3\arcmin$), is imaged with a target acquisition camera (type AP6, from
Apogee Instruments). We use the following nomenclature for the fibre channels:
channel 0 for the central fibre, channels 1 to 6 for the ring fibres and
channel 7 for a sky background fibre, located about $1\arcmin$ off-set from the
target fibre. For the observations we used tapered fibres with $320~\rm{\mu m}$
diameter at the pick-up  and an exit diameter at the detector of
$100~\rm{\mu m}$.  The single fibre size  in the focal plane is  equivalent to
a $2\farcs344$ resolution at the 2.56-m NOT telescope. Although the pure silica
core fibres transmit light from 400~nm  to $\sim $950~nm with better than
$99\%$, the net transmission of  the tapered fibres is lower and appears to
range between $60\%$ and $90\%$. 

The  timing of individual  photons is  controlled by  signals from  the Global
Positioning  System (GPS)  to an  absolute  accuracy of  $\sim 2~\rm{\mu  s}$,
although the readout system limits  the resolution to $\sim 4~\rm{\mu s}$. The
OPTIMA detector is operated with two PCs and is autonomous except for the need
to have a good telescope guiding system.

\subsection{Rotating Polarisation Filter}
\label{Sec:RPF}
For the present observations OPTIMA was equipped  with a  rotating  polarisation
filter (RPF) in the incoming beam above the fibre pick-up so that  all fibre
channels and the CCD image are fully  covered. The  polarising filter (Type
10K by Spindler \& Hoyer) is mounted on a precision roller bearing and is
rotated with typical frequencies of a few Hz (the averaged frequency over
  the whole  observations was 3.4~Hz). Incoming linearly polarised  light is
then modulated  at twice the rotation  frequency of the  filter. The reference
position  of the filter  is given  by a  signal from  a magnetic  switch (Hall
sensor) which is  registered and timed in  the same way as a  photon event and
stored in a 
separate DAQ channel. The position of the polarising filter for any photon event
is then derived by interpolating the photon arrival time between the preceding
and the following Hall sensor signal, i.e.:
\begin{displaymath}
\alpha_{\rm RPF}(TOA)= \frac{t_{TOA} - t_b}{t_a  - t_b} \times 360\degr,
\end{displaymath}
where $\alpha_{\rm RPF}(TOA)$  is the RPF angle  at which the  event was observed,
$t_{TOA}$  is the  event time  of  arrival, and $t_b$ and $t_a$ are  the
recorded  times  of  the  Hall  signal  sensor  before  and after  $t_{TOA}$.
Since, during  one turn of the RPF all possible polarisation angles are measured
two times,  for all values bigger than  $180\degr$ exactly $180\degr$ was
subtracted.

Slight irregularities in the rotation frequency of  the RPF that  occur on
time-scales  longer than fractions of  a second,  e.g.  due to  supply voltage
drifts or  mechanical resistance changes in the bearing  and motor, can thus be
corrected with sufficient accuracy.  The  RPF  was tested in the  lab  with
unpolarised and  linearly polarised light to ensure and prove  that the OPTIMA
fibres and detectors have no  intrinsic systematic response to polarised  light
\citep{Kellner2002}.

The polarising  filter modulates  the incoming light  effectively only  over a
wavelength  range of  about 470-750~nm.  Since the  APD response  (QE$>20\%$)
extends  from  about 450~nm  to  970~nm and  no  wavelength  information of  the
individual recorded  events is available,  it is necessary to  block radiation
outside the filter modulation range. Such photons, especially towards the near
IR,  are not  modulated and  would  decrease the  estimate for  the degree  of
polarisation. Therefore, we inserted  an  IR blocking  filter  that cuts  the
wavelength range at about 750~nm.

The method to derive the polarisation characteristics follows the approach
described by \citet{Sparks1999}.  Details are given in Appendix  \ref{Sec:Appendix}
to this paper.

\section{Observations}
The Crab observations were performed on November 23--27 2003 (52966 -- 52971
MJD). They consists of 160 pointings of ten minutes each. About one third of
the total number of counts falls in the central fibre, whereas around 5 to 10
percent are registered in each of the background fibre (channels marked from 1
to 6 in the Fig. \ref{Fig:HST}). The count rate for the central fibre decreases
when the seeing deteriorates and photons spill out into the ring channels
($1\div6$). In order to screen for observation intervals of acceptable seeing we
therefore apply a cut on the fraction of total counts in channel 0 at the level
of $30\%$. After screening for good seeing and proper pointing, our data base
resulted in 83 files, equivalent to 13 hours and  43 minutes of total exposure.

\section{Data reduction}
\subsection{Flat field correction}

\begin{table}
\centering
\caption{Fibre flat field correction factors}
\label{Tab:flat_field}
\begin{tabular}{@{}c c}
\hline\hline
Channel & Factor \\
\hline
0 & $1.000 \pm 0.012 $ \\
1 & $1.009 \pm 0.013 $ \\
2 & $1.074 \pm 0.013 $ \\
3 & $0.899 \pm 0.012 $ \\
4 & $0.850 \pm 0.011 $ \\
5 & $0.956 \pm 0.012 $ \\
6 & $1.140 \pm 0.014 $ \\
\hline
\end{tabular}
\end{table}

For flat field correction factors of all channels we used two sets of dark
sky observations, i.e.  pointing  at $\alpha=  00^{h}  07^{m}  02\fs00$,
$\delta = +73\degr 03\arcmin 28\farcs00$. Both observations were taken on
November 25. The data acquisition started at 21:39:10 UTC and  21:49:13 UTC.
The exposure time amounted to 600 and 300 seconds for the first and second
observation, respectively. The detected count rates behind the filters
fluctuated from 180~Hz for channel~4 up to 240~Hz for channel~6. To obtain
the flat field correction factors  we assumed that this value was 1.0
for the central fibre, and we scaled the count rates of the other fibres
to the central one. We binned the data in  ten seconds intervals,
and then calculated the average and standard deviation for each single
channel  of the ring fibres. The resulting numbers are shown in
Tab.~\ref{Tab:flat_field}.

\subsection{The HST polarisation standards}

\begin{table}
\centering
\caption{HST polarisation and photometric standards chosen from
the list of \citet{Turnshek1990}}
\label{Tab:pol_stand}
\begin{tabular}{@{}l c r}
\hline\hline
Name / & Coordinates FK5 & Comments \\
Spectral Type & & \\
\hline
\multicolumn{3}{c}{\bf HST Polarisation Standards} \\
 & & \\
{\bf BD+64 106} & $~00^{h}~57^{m}~36\fs70$         & V = 10.34 \\
B1V             &  $+64\degr~51\arcmin~34\farcs9$  & $p = 5.65\% \pm 0.053\%$ \\
& & $\theta$ = $96\fdg8$ \\
 & & \\
{\bf G191B2B} & $~05^{h}~05^{m}~30\fs61$  & V = 11.79 \\
WD                & $+52\degr~49\arcmin~51\farcs9$ & $p = 0.09\% \pm 0.048 \%$
\\
 & & \\
\multicolumn{3}{c}{\bf HST Photometric Standard} \\
 & & \\
{\bf GD50}      & $~03^{h}~48^{m}~50\fs20$            & V = 14.06 \\
WD              & $-00\degr~58\arcmin~31\farcs2$      &  \\
\hline
\end{tabular}
\end{table}

\begin{table*}
\centering
\caption{Measurements of  the HST  polarisation standard BD+64~104. Date and
time of observations, as well as exposure length in seconds are given in the
first, second and third row, respectively. Obtained values of P.D., $p$,
(upper value) and P.A. ($\theta$, bottom value) are given for each
OPTIMA detector channel from 0 to 6 ($p$ and $\theta$ are in percent and
degrees, respectively). Additionally, last  row (assigned as `$1\div6$' )
gives $p$ and $\theta$ values obtained after averaging the Stokes  parameters
from all ring channels. \emph{Mean} gives $p$ and $\theta$, as well as their
standard deviations, after averaging all existing data of BD+64~104.
The obtained position angle of BD+64~104 (bold faced text in the row of
channel 0) was later used for the OPTIMA instrument calibration.}
\label{Tab:hst_ps}
\begin{tabular}{@{}lccccccccc}
\hline
\multicolumn{10}{c}{\textbf{BD+64~104}} \\
\hline
\hline
Obs. Date &  & \multicolumn{2}{c}{25 Nov 2003} &  & \multicolumn{3}{c}{27 Nov 2003}
 & & Mean\\
Time [UTC] & & 22:07:01 & 22:09:46 & & 21:33:35 & 21:36:08 & 21:38:34 & & \\
Expo. [s] &  & 140 & 130 & &  110 & 120 & 120 & & \\
\hline
Channel & & & & & & & & & \\
0  & & 1.30 & 1.35 & & 1.98 & 1.86 & 1.67 & & $1.63\pm0.30$\\
   & &    \textbf{96.6} &
          \textbf{96.8} & &
          \textbf{96.2} &
          \textbf{96.9} &
          \textbf{97.2} & &
          \textbf{96.74$\pm$0.37} \\
& & & & & & & & & \\
1 & & 4.57 & 4.64 & & 3.71 & 3.60 & 4.21 & & $4.15\pm0.48$\\
  & & 100.3 & 96.6 & & 98.0 & 96.5 & 98.3 & & $98.14\pm1.62$\\
& & & & & & & & & \\
2 & & 5.16 & 5.17 & & 4.59 & 4.84 & 5.57 & & $5.07\pm0.37$\\
  & & 96.0 & 94.7 & & 98.0 & 96.8 & 96.8 & & $96.46\pm1.22$\\
& & & & & & & & & \\
3 & & 5.38 & 5.16 & & 5.49 & 5.08 & 5.04 & & $5.23\pm0.20$\\
  & & 94.3 & 92.9 & & 95.5 & 95.9 & 97.4 & & $95.20\pm1.70$\\
& & & & & & & & & \\
4 & & 4.67 & 4.34 & & 4.14 & 4.98 & 4.27 & & $4.48\pm0.34$\\
  & & 99.1 & 98.6 & & 98.5 & 100.5 & 103.5 & & $100.04\pm2.09$\\
& & & & & & & & & \\
5 & & 5.51 & 5.26 & & 5.53 & 5.58 & 5.38 & & $5.45\pm0.13$\\
  & & 96.7 & 94.3 & & 97.9 & 96.2 & 97.1 & & $96.44\pm1.35$\\
& & & & & & & & & \\
6 & & 4.62 & 5.00 & & 5.06 & 4.52 & 4.80 & & $4.80\pm0.23$\\
  & & 98.6 & 98.7 & & 93.6 & 93.6 & 91.7 & & $95.24\pm3.21$\\
& & & & & & & & & \\
1$\div$6 & & 5.06 & 4.95 & & 4.47 & 4.37 & 4.73 & & $4.72\pm0.30$\\
  & & 96.7 & 95.1 & & 96.8 & 95.8 & 96.4 & & $96.16\pm0.71$\\
\hline
\end{tabular}
\end{table*}

\begin{table*}
\centering
\caption{Measurements of  the HST  polarisation standard G191B2B and the HST
photometric standard GD50. Date and time of observations, as well as exposure
length in seconds are given in the first, second and third row, respectively.
Obtained values of P.D., $p$, are given in percent for each OPTIMA
detector channel from 0 to 6. Additionally, last  row (assigned as `$1\div6$')
gives $p$ values obtained after averaging the Stokes parameters from all ring
channels. \textit{Mean} gives $p$ and standard deviations after averaging all
existing data of G191B2B from Nov 26th. In the last column the results of GD50
measurements are given.}
\label{Tab:hst_ups}
\begin{tabular}{@{}lcccccccccccc}
\hline
\hline
\multicolumn{9}{c}{\textbf{G191B2B}} & & \multicolumn{1}{c}{\textbf{GD50}} \\
\hline
Obs. Date &  & \multicolumn{4}{c}{26 Nov 2003} &  & Mean &  & & 27 Nov 2003 \\
Time [UTC] & & 21:33:08 & 21:36:22 & 21:41:01 & 21:43:53 & & & & & 21:52:17 \\
Expo. [s] & & 160 & 150 & 130 & 130 &  &  &  & & 590 \\
\hline
Channel & & & & & & & & \\
& & & & & & & & \\
0 &  & 0.04 & 0.07 & 0.03 & 0.06 & & \textbf{0.05 $\pm$ 0.02} & & & 0.08 \\
1 &  & 0.37 & 0.48 & 0.47 & 0.21 & & 0.38 $\pm$ 0.13 & & & 0.30 \\
2 &  & 0.75 & 0.41 & 0.65 & 0.85 & & 0.67 $\pm$ 0.19 & & & 0.44 \\
3 &  & 0.84 & 0.46 & 0.80 & 0.72 & & 0.71 $\pm$ 0.17 & & & 0.61 \\
4 &  & 0.30 & 0.66 & 0.69 & 0.47 & & 0.53 $\pm$ 0.18 & & & 0.62 \\
5 &  & 0.16 & 0.56 & 0.36 & 0.16 & & 0.31 $\pm$ 0.19 & & & 0.23 \\
6 &  & 0.14 & 0.21 & 0.04 & 0.42 & & 0.20 $\pm$ 0.16 & & & 0.53 \\
& & & & & & & & \\
1$\div$6  &  & 0.35 & 0.34 & 0.30 & 0.32 & & 0.33 $\pm$ 0.02 & & & 0.22\\
\hline
\end{tabular}
\end{table*}

We  performed observations of stars with well known polarisation in order to
understand  the intrinsic  polarisation and  the response  of the instrument. 
Since OPTIMA is very sensitive and limited  in its capacity for data acquisition
at high  rates (count rates above 37~kHz lead to noticeable, but correctable,
pile-up effects, \citealt{Muehlegger2006}) we selected three of the weakest
stars from the list of \citet{Turnshek1990}; two polarisation and one
photometric standards (Tab.~\ref {Tab:pol_stand}). The  first star of the
two polarisation standards is highly polarised, on  the level of $6\%$,
whereas the second one has a very low polarisation degree of $\sim$1
\textperthousand . Unfortunately for our measurements both of them are
quite bright by OPTIMA standards. Therefore, we additionally performed optical
polarisation measurements of a dimmer star, i.e. a photometric standard from
\citet{Turnshek1990} list, expecting its light not to be polarised.

We observed  the HST polarisation standard BD+64~106 for 270 and 350 seconds
on the 25th and 27th of November, respectively (Tab.~\ref{Tab:hst_ps}).
The observing conditions during the two exposures were very different,
requiring a detailed discussion of the results. On Nov 25, 22:06 -
22:12 UTC the average seeing was on the level of $1\farcs27$ (RoboDIMM
measurements), whereas for  the time span 21:33 - 21:40 UTC on Nov 27 no
information about the  seeing conditions  was available from the
RoboDIMM telescope.  At the beginning of this night the weather  was very good.
But later, during less than two hours, between 20:00 and 22:00 hours, the humidity
changed from 20\% to almost 60\%, and the wind speed increased up to 15~\rm{m/s}.
Therefore, the seeing on Nov 27 was probably worse than during Nov 25,  likely
more  than $2\farcs0$. This difference in seeing causes different count rates
in the central and ring fibres. On 25 Nov  the count rate in channel~0 was about
190~kHz, and in the ring fibres $\approx 20~\rm{kHz}$, while during Nov 27 we had 
150~kHz (centre) and $\approx 50~\rm{kHz}$ (ring). As a consequence of the high count
rates ($\gg 37~\rm{kHz}$) the central channel
was severely affected by pile-up in these observations. For the second
set of data (27 Nov) it happened that some of the ring fibres, especially
channels 1 and 6, were also saturated. The values of the polarisation degree and
the position angle for all channels and for all sets  of observations are
gathered in Tab.~\ref{Tab:hst_ps}. It is important to notice that, even if
the central fibre was suffering pile-up, the calculated position angle for
each single observation  of BD+64~104 is very  much constant. This is caused
by the fact that the position  angle, being the phase of the maximum of the
modulated incoming light,  is not strongly dependent on the pile-up effect.
Saturation is however important when the amplitude of RPF modulation,
corresponding to the degree of polarisation, is to be measured. We can  observe 
this by comparing results in Tab.~\ref{Tab:hst_ps}. For the central fibre  there
is significant difference in  the polarisation degree for  the first and second
data set. When  the dispersion of the stellar image due to seeing was larger
the star light was more smoothly  distributed among  all channels, therefore an
increase of $p$  in the central fibre is observed. Seeing conditions, being
equivalent  to the count rates in the ring, did not affect significantly  the
measured values  of  $p$  in the ring fibres. Only in the case of  channel~1
the polarisation degree is much smaller during the Nov 27 observations than
Nov 25.  This is strongly connected with the fact that, during  the second
observation, this channel, among all of the ring channels, was the most
saturated one.

Taking into account the resulting $\theta$ and $p$ (Tab.~\ref{Tab:hst_ps}) of
the  HST polarisation standard  BD+64~104 (being  a very  bright star as for the
OPTIMA instrument) we can conclude that for the purpose of calibrating the north
direction  of our instrument it  is  safe  and valid  to use  values  of the
polarisation angle from the  central fibre (Tab.~\ref{Tab:hst_ps}, bold faced
text). The averaged value of $\theta_{\rm{BD+64  104}}$ amounts to $96.74\degr
\pm0.37\degr$.  This result was obtained by shifting the intrinsic RPF angles
(the position of the Hall sensor indicates the origin of the intrinsic angles)
by $92\degr$. The angles  $\theta$ given  here  and hereafter  in
the text  and tables are the  $E-$vector position  angles  relative to celestial
north (N to E). Due to the pile-up in the central fibre the best estimate of the
polarisation degree of BD+64~104 comes from the sum of the light detected in the
ring fibres. The obtained  value, $4.72\%\pm0.30\%$, is somewhat smaller than 
the value given by \citet{Turnshek1990}, $5.65\%\pm0.053\%$, but one should take
into account the systematic problems encountered with such a bright standard
star. In this case the bad seeing  (high value) is an advantage, because it
naturally defocuses our target. We also performed  observations of the
unpolarised HST  standard G191B2B (Tab. \ref{Tab:pol_stand} and
\ref{Tab:hst_ups}). Data for this target were taken on 26 Nov, 21:33 UTC for
about 10 minutes with seeing  being on the level of $1\farcs0$. We found a very
small polarisation degree for this star, as expected, and thus the position
angle has little meaning. During these observations the count rate was on the
level  of 150~kHz and  $\la 40~\rm{kHz}$  in the central  and  ring
fibres, respectively. We should point out that channel 0 was affected by
pile-up and the result $0.05\%\pm0.02\%$ (Tab.~\ref{Tab:hst_ups})
might be biased, even if it is in good agreement with the value
$0.09\%\pm0.048\%$ given by \citet{Turnshek1990}. The averaged
value of P.D. from the background fibres is higher.  It amounts to
$0.33\%\pm0.02\%$, and is  probably  partly  contaminated  by   the
polarised sky background.

For  the  purpose  of  the  instrument  calibration we  also  chose  from
the \citet{Turnshek1990} atlas of HST photometric, spectroscopic, and
polarimetric calibration objects a star with brightness that  should   assure
count rates lower that the pile-up threshold, i.e. GD50
(Tab.~\ref{Tab:pol_stand}). It is a photometric standard, thus we expect its
light not to  be polarised. Observations of this target were  performed on
Nov 27th at 21:52 UTC with recorded  count  rates   in  the  central  channel of
10-15~kHz and in the ring channels $\la 1~\rm{kHz}$. Results are shown in the
Tab.~\ref{Tab:hst_ups}. The source had polarisation degree below 
1\textperthousand ~(measured in the central fibre), whereas  the measured
$p$ within  the  ring fibres  was about three time greater. These values
correspond to the sky background polarisation. However, the sky background could
have been more highly polarised. The contribution of the night sky to the
observed count rates was on  the level of one quarter in each fibre only, i.e.
the background ring fibres were contaminated by the source. Thus, the signal
measured in the ring fibres was most probably depolarised and could have higher
polarisation degree in reality.

\subsection{Raw data binning for the Crab analysis}
\label{Sec:rdb}

\begin{table}
\centering
\caption{The Crab pulsar radio ephemeris}
\label{Tab:ephemeris}
\begin{tabular}{@{}l c}
\hline
\hline
Parameter  & {\rm Value} \\
\hline
R.A. (J2000), $\alpha_{2000}$ & $05^{h} 34^{m} 31\fs972$ \\
DEC (J2000), $\delta_{2000}$ & $+22\degr 00\arcmin 52\farcs07$ \\
\rm{Valid range (MJD)} & 52944--52975\\
Epoch, $t_0$ \rm{(TDB MJD)} & 52960.000000296\\
$\nu_0$ \rm{(Hz)} & 29.8003951530036\\
$\dot{\nu_0}~\rm{(10^{-10}~Hz~s^{-1})}$ & -3.73414\\
$\ddot{\nu_0}~\rm{(10^{-20}~Hz~s^{-2})}$ & 1.18 \\
\hline
\end{tabular}
\end{table}

\begin{figure}
\centering
\includegraphics[scale=0.5]{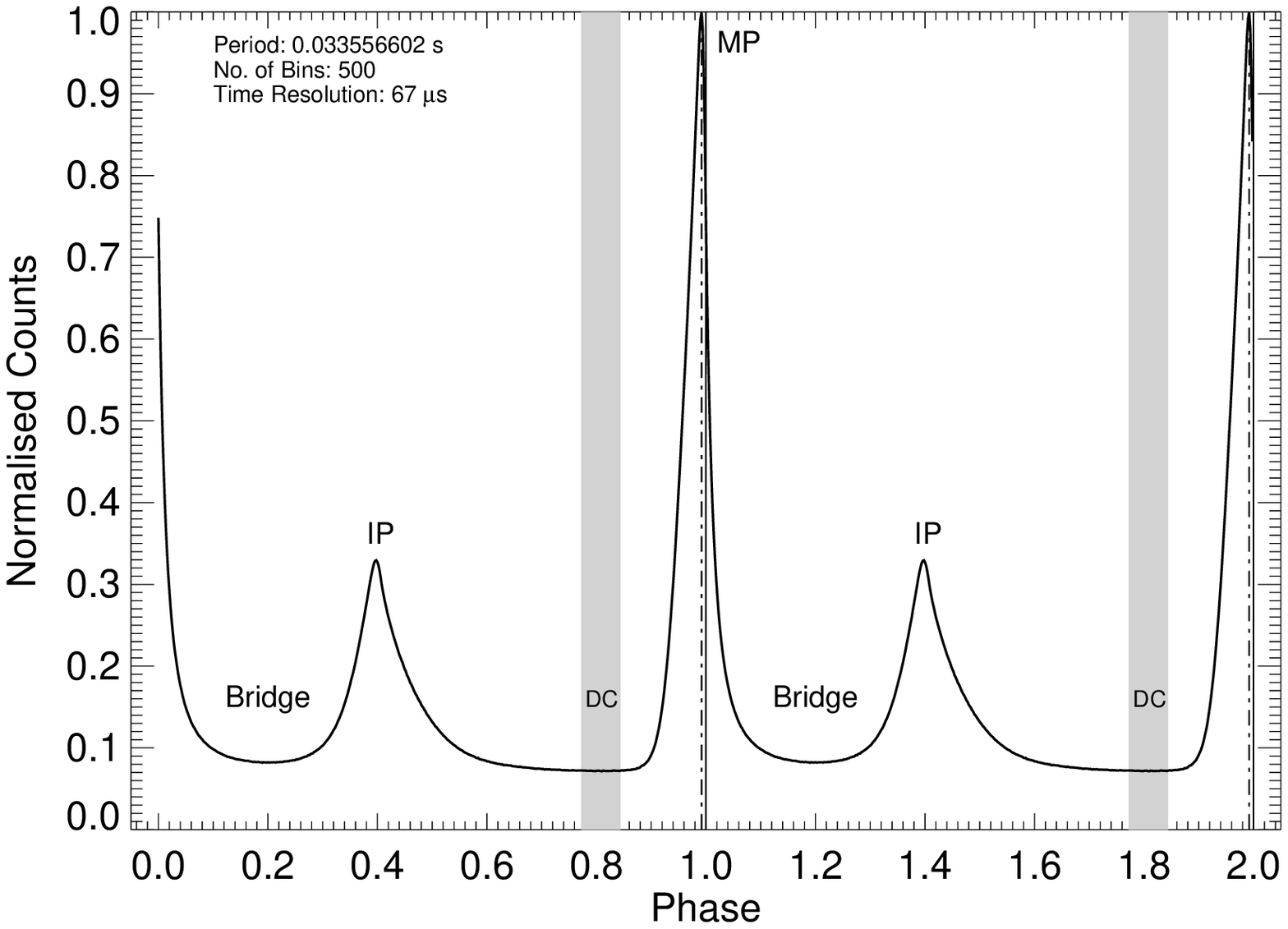} \\
\includegraphics[scale=0.5]{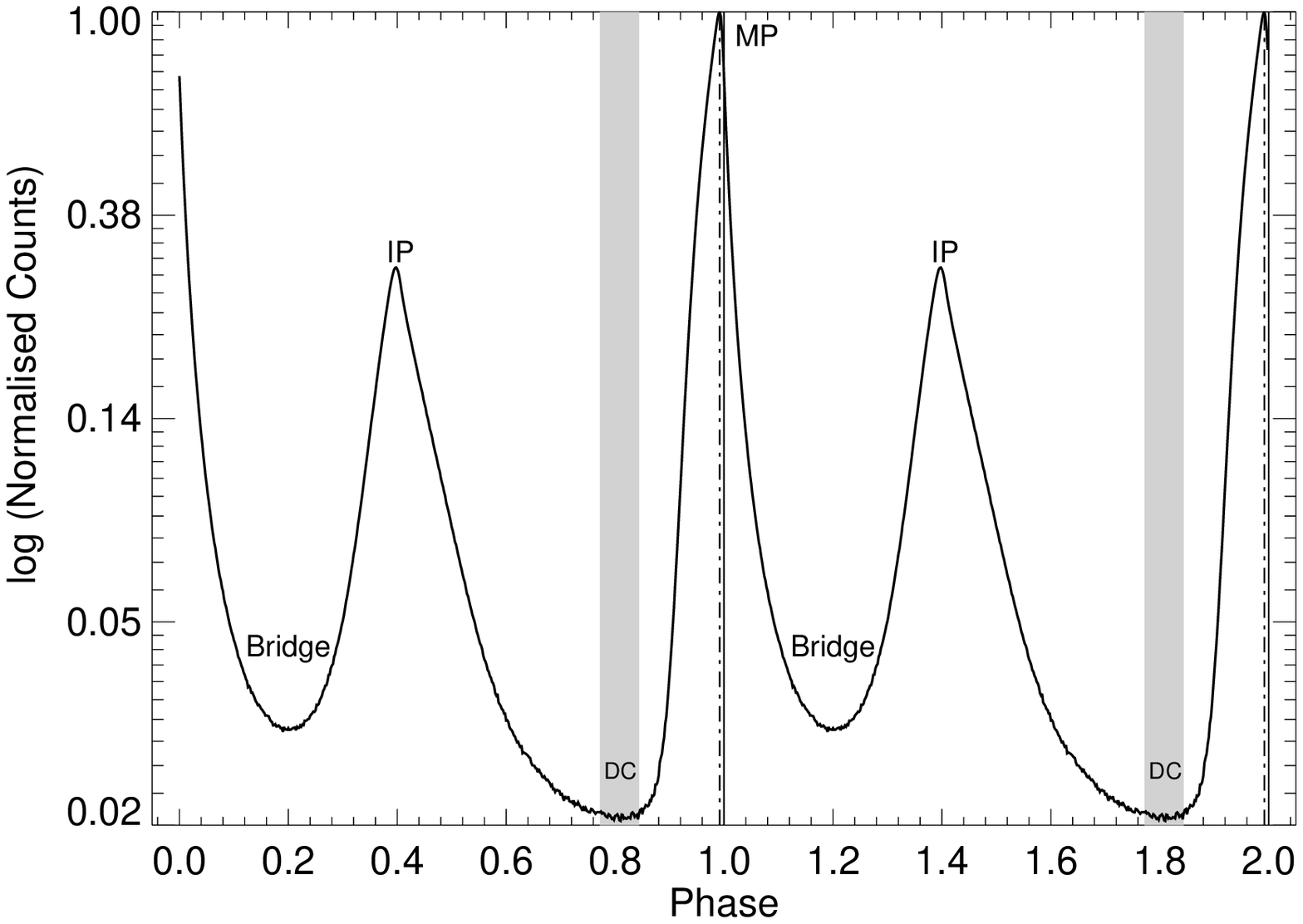}
\caption{The  Crab pulsar  light  curve  obtained from  photons recorded by
the OPTIMA central fibre APD (channel 0). The components of the light curve
are indicated  as follows:  main pulse  MP, inter  pulse IP,  non-zero intensity
level between MP and IP, i.e. bridge, as well as the DC region, previously known
as the `off-pulse' component in   in  the phase range $0.7729 - 0.8446$ . The upper
panel shows the raw light curve including counts from the nebula and the sky background.
In the lower panel the lightcurve is shown in a logarithmic scale after subtraction
of the nebular and sky backgrounds. This display shows clearly the DC intensity level.
Two rotation periods are shown  for clarity.}
\label{Fig:raw_lc}
\end{figure}

For  each incoming  and detected  photon  the OPTIMA  data acquisition  system
stores its Time Of  Arrival (TOA) in a compact proprietary binary format with a
resolution of $4~\rm{\mu s}$. By using  the  OPTIMA system  software
\citep{Straubmeier01a} one can obtain TOAs  in units of Julian Date (JD). 
In  order to apply the rotational model of the pulsar derived from radio
observations  to our optical photons, it is required to measure the TOAs
in an `inertial observer frame'. Using the NOT  position coordinates, the
Crab  pulsar coordinates (Tab. \ref{Tab:ephemeris}) and as  planetary ephemeris
the JPL  DE200 tabulations \citep{Standish1982} we  transformed the recorded
TOAs to the commonly used, and (for these purposes) inertial frame, i.e. to
TOAs at the solar system barycentre.

Parameters    of   the   rotational   model   of    the   Crab   pulsar
(Tab.~\ref{Tab:ephemeris})   obtained  from   the  radio   observations  are
regularly, once per month, published  by the Jodrell Bank Observatory pulsar
group  in the Crab  Pulsar  Monthly  Ephemeris\footnote{
\texttt{http://www.jb.man.ac.uk/$\sim$pulsar/crab.html}} \citep{Lyne1993}.
For each  recorded event (TOA) the corresponding pulsar phase and the phase
of the RPF is calculated. We sort  the individual events  into a 3D  data array.
The first  dimension is given  by  the  rotational phase  of  the  pulsar (binned
with various resolutions), the  second one by  the phase of the polarisation filter
at  each photon TOA (binned in $1\degr$ intervals), and the third dimension is the
fibre number or number of DAQ channels (seven channels). 

To obtain an 'unpolarised' light curve, e.g. with a time resolution of
$67~\rm{\mu s}$ which assures a good S/N ratio even for the phase ranges where
the intensity is very low, the data array is constructed with 500 bins per
rotational period of the Crab, and the events spread out over the RPF phases
are all added up. The Crab light curve, in its raw form as the measured count
rate in the central fibre without background subtraction, is shown in the top
panel of Fig.~\ref{Fig:raw_lc}. We indicate the components main pulse (MP),
inter pulse (IP), non-zero intensity level between two peaks - bridge, and the
DC region, previously known as the so called `off-pulse' component. For our
further analysis we define the DC component as the counts between phases
$0.7729 - 0.846$, which is in accordance with the findings of
\citet{Percival1993}. The same light curve after background subtraction
(method described in the next section) is shown in the lower panel of
Fig.~\ref{Fig:raw_lc} with a logarithmic scaling to enhance the visibility of
the DC level.  The higher S/N ratio in the peaks allows us to use better time
resolution for these phase ranges, i.e. up to 3000 bins per period, which
corresponds to a bin of 11~\rm{$\mu s$} interval (Figs.~\ref{Fig:pa_pd},
\ref{Fig:zoom}, \ref{Fig:zoom_dc}).

\section{Results for the Crab nebula and pulsar}
\subsection{Nebular contribution}

\begin{table}
\centering
\caption{The Crab nebula's P.D. and P.A. at the DC phase range}
\label{Tab:neb_res}
\begin{tabular}{lcc}
\hline\hline
\hline\hline
Channel & $p$ [\%] & $\theta$ [$\degr$] \\
\hline
0 & 33.08(25) & 118.8(0.2) \\
1 & 9.05(8) & 141.3(2) \\
2 & 10.30(8) & 142.5(2) \\
3 & 8.77(7) & 148.4(2)  \\
4 & 9.35(8) & 136.6(2)  \\
5 & 11.50(8) & 133.1(2) \\
6 & 10.26(7) & 137.9(2) \\
1$\div$6 &  9.71(8) & 139.8(2)\\
\hline
\end{tabular}
\end{table}

\begin{figure}
\includegraphics[scale=0.45]{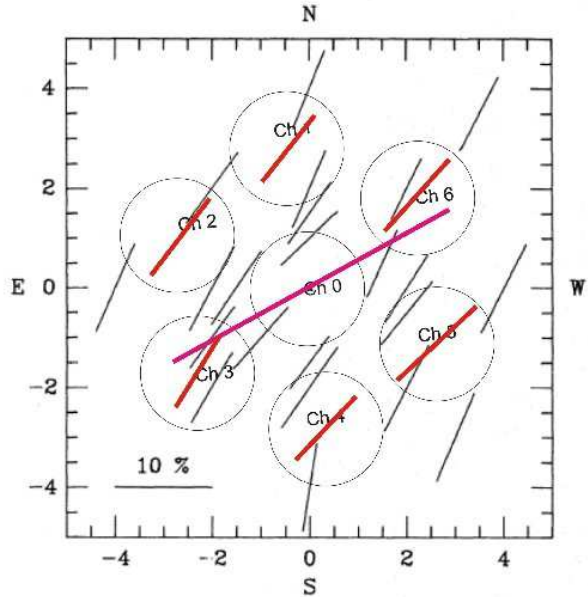}
\caption{Polarisation of the Crab nebula as measured by each single fibre at the minimum
phase of the pulsar light curve. It is graphic representation of results given in Tab. \ref{Tab:neb_res}. The axes are calibrated  in arc seconds centred  on  the pulsar. The scaled OPTIMA fibre bundle is over  plotted on the figure given in \citealt{Smith1988}. The aperture of a single fibre is $2\farcs35$. The pink line for central fibre and the red lines for the ring fibres represent our measurements with the rotating polaroid filter. The black lines are the results that \citealt{Smith1988} determined with a polarimeter based on a rotating half-wave plate and a Foster prism.}
\label{Fig:Smith}
\end{figure}

The Crab synchrotron nebula is  a relativistic magnetised plasma that is powered
by  the spin-down energy of the pulsar. It was the first recognised astronomical
source  of   synchrotron radiation. The synchrotron nature of the radiation was
confirmed by optical polarisation observations \citep{Woltjer1957}. The
conversion efficiency of the nebula is quite high, with 10\%--20\% of the
spin-down  energy  released by  the pulsar appearing as synchrotron radiation.
The inner synchrotron nebula is a region consisting of jets, a torus of X-ray
emission, small-scale variations in polarisation and spectral  index,  and
complexes  of  sharp wisps. Most theoretical models associate the
sharp wisps  seen at visible and radio wavelengths with the  location of the
shock wave between  the pulsar and the synchrotron nebula. Closer  to  the outer
boundary of the nebula are the  filaments - the  chemically enriched material
ejected during the  supernova explosion  observed by  Chinese astronomers in
1054.

For a long time the imaging  investigations of the Crab nebula were limited in a
fundamental  way by the  spatial resolution of  the detectors, not being able to
reveal a  wealth of  subarcseconds structures. A  breakthrough in  the optical
studies of the  structure of the Crab nebula was undertaken  by using the Wide
Field  and Planetary  Camera  2 (WFPC2)  on  board of  the \emph{Hubble  Space
Telescope}. \citet{Hester1995}  performed   this   observation  with $0\farcs1$
resolution. Below we shortly describe the  most important  HST results that are
important in  the context  of this paper.

They discovered  a  bright  knot of  visible  emission located
$0\farcs65$ to  the south-east  of the  pulsar, along the  axis of the system
(Fig. \ref{Fig:HST}). This  inner  knot,  along with   a  second  similarly
sharp but  fainter  knot (hereafter outer  knot) located at a  distance of
$3\farcs8$  from the pulsar, lies at an  approximate position angle of
$\sim 115\degr$  east to north. Both knots  are aligned with  the X-ray
and optical  jet to  the south-east  of the pulsar and  are elongated  in
the dimension  roughly perpendicular to  the jet direction, with lengths
of about a half arc second. Both,  the inner and outer knot appear to be present but
not well resolved  in the images of the Crab nebula previously taken by ground
based telescopes. Fig.~\ref{Fig:HST} is an enlargement of the co-added HST WFPC2
images of 12 observations of the Crab nebula in between 2000 and 2001 with the
F547M filter \citep[observations group index 6 of][it was kindly
supplied by  Roger Romani, private  communication]{Ng2006}. The pulsar is
identified with the lower/right of the two stars near the geometric centre of
the nebula. The OPTIMA fibre bundle  centred on  the pulsar and scaled with the
NOT focal plane scale is over plotted.  It is clear that we are not able to resolve
the Crab pulsar from the inner  knot within the central fibre. Moreover, with
seeing being bigger than $1\farcs0$ the pulsar light is somewhat spread into
the ring channels. Evidence  of this effect is seen in the light curves when
the  intensities measured  by individual ring fibres are folded with the pulsar
phase and pulsed emission is detected.  Additionally, there may be contributions
of photons coming from the outer knot (not very clearly seen in the
Fig.~\ref{Fig:HST}), but an indication of it can be seen  on the south-east
side in between the fibres 3 and 4. It  is also noteworthy to mention that
the ring fibres see different patches of nebulosity. One third of all
counts are recorded in the central fibre. The background fibres contribute to
the total number of counts in the percentage of 9, 10, 14, 11, 12, 11 for channels
from 1 to 6, respectively.

To  obtain the  polarisation characteristics  of the  pulsar neighbourhood
we assumed that within the DC  phase range  the contribution  of the pulsar
emission to  the ring fibres is minimal. The integrated pulsed contribution
in each fibre (spill-out from the pulsar) with respect to the total counts
in the fibre is on the level of $3.1\%$, $4.5\%$, $4.4\%$, $2.5\%$, $2\%$,
$1.9\%$ for the channels from 1 to 6, respectively. Therefore, during our
background (nebula) calculations we consider only light coming  within the
`off-pulse' phase range, i.e. 7\% of the whole rotational cycle of the Crab
pulsar. Obtained polarisation degree and position angles for  each of the single
OPTIMA  apertures  are given in Tab.~\ref{Tab:neb_res}, as well as illustrated
in the Fig. \ref{Fig:Smith}. We compared our results with  previous
ones by over plotting them on the polarisation sky map of the very close
neighbourhood presented by \citet{Smith1988}. By averaging the  Stokes parameters
over all background channels we get $p= 9.71\%\pm0.08\%$ and
$\theta= 139.8\degr\pm0.2\degr$ for the region surrounding the pulsar. Close to
the pulsar the nebular polarisation is  quite uniform ($\sim 9-11\%$) but the
position angles change  steadily with radial  distance. Two to three arc seconds
from the pulsar the mean value is around $140\degr$ but beyond five arc seconds
the position angle exceeds $155\degr$  and  it  is very position-dependent
\citep{McLean1983}.

\subsection{Polarisation characteristics of the Crab pulsar}

\begin{figure*}
\includegraphics[scale= 0.5]{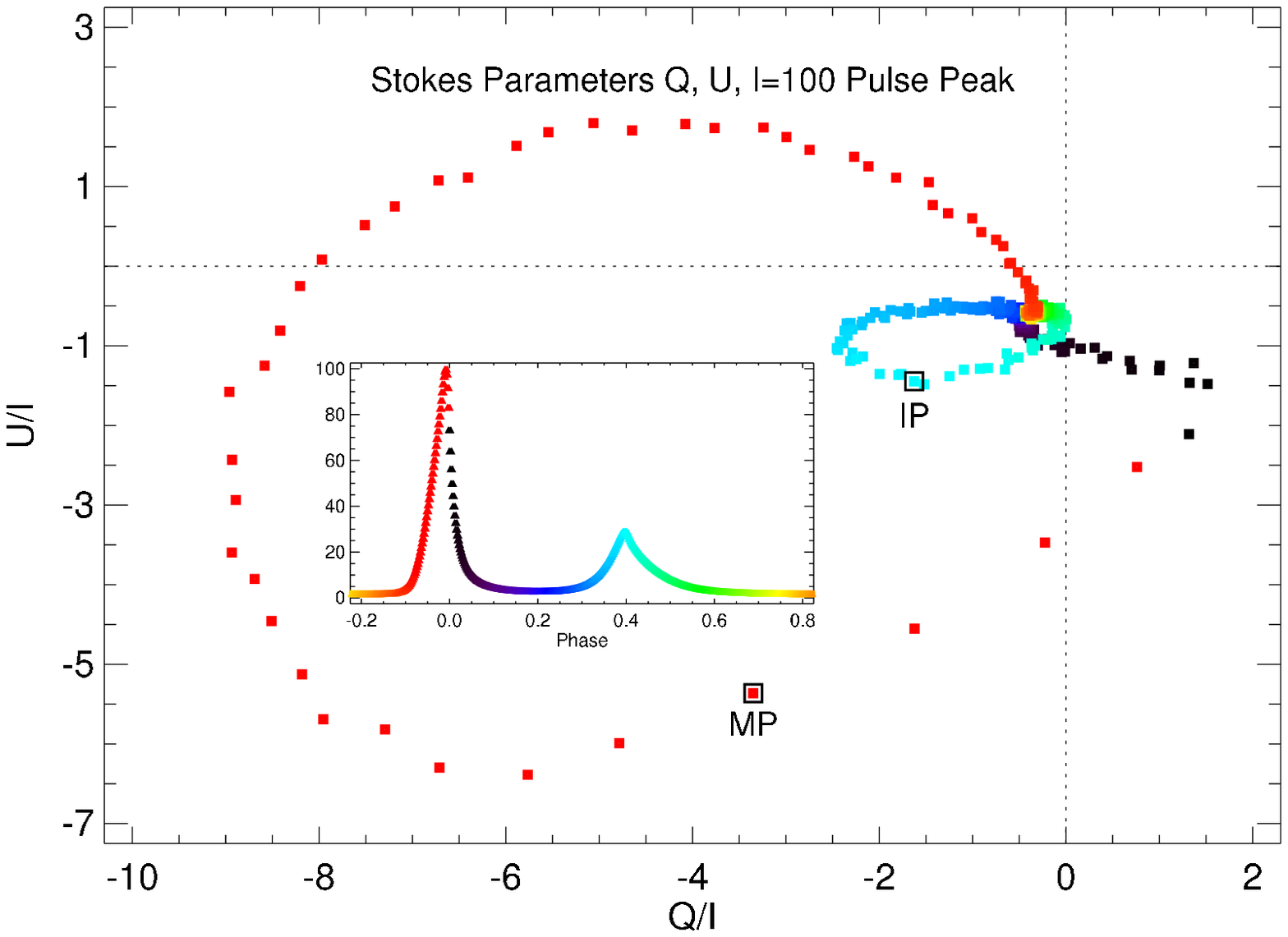}
\includegraphics[scale= 0.5]{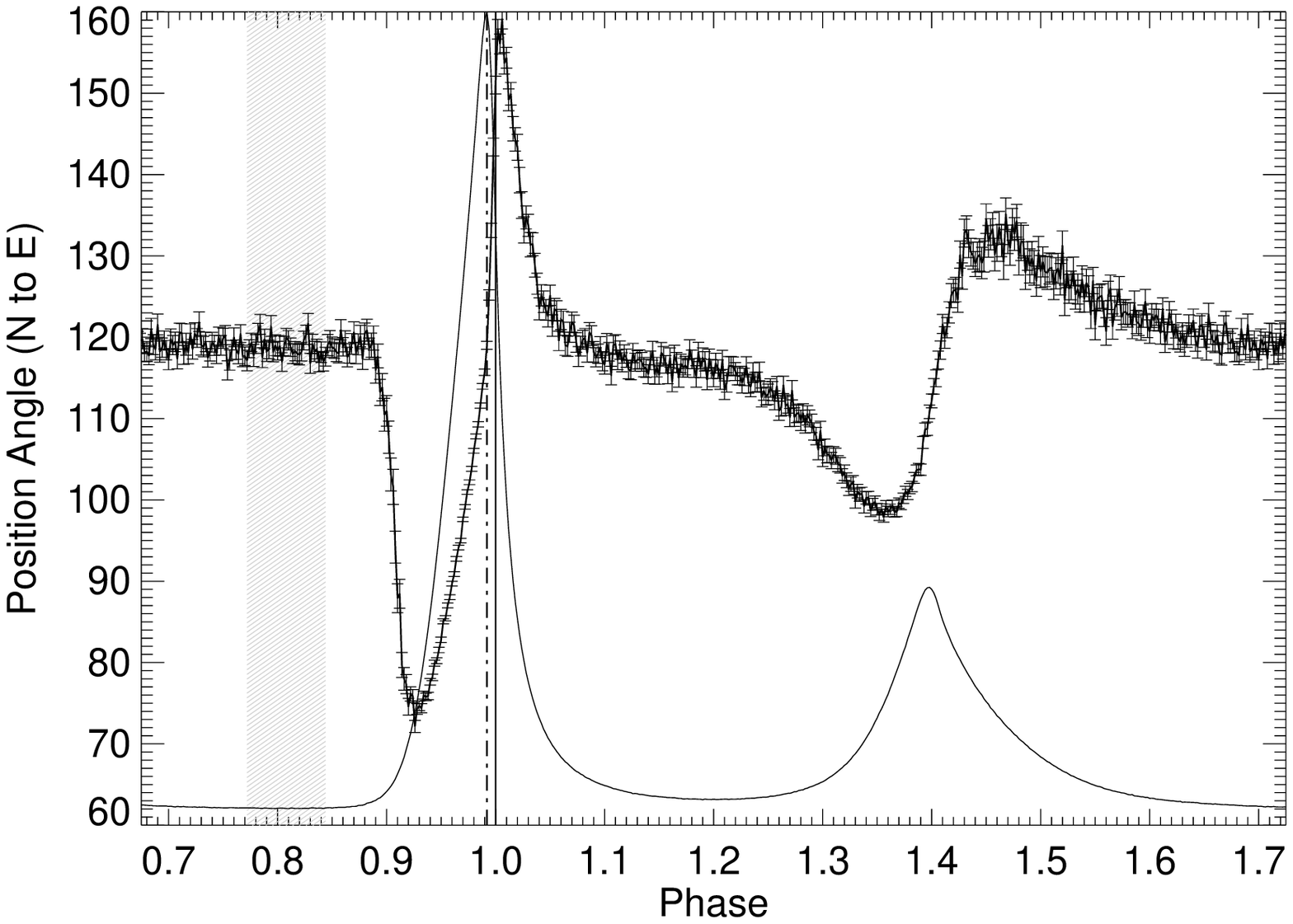}\\
\includegraphics[scale= 0.5]{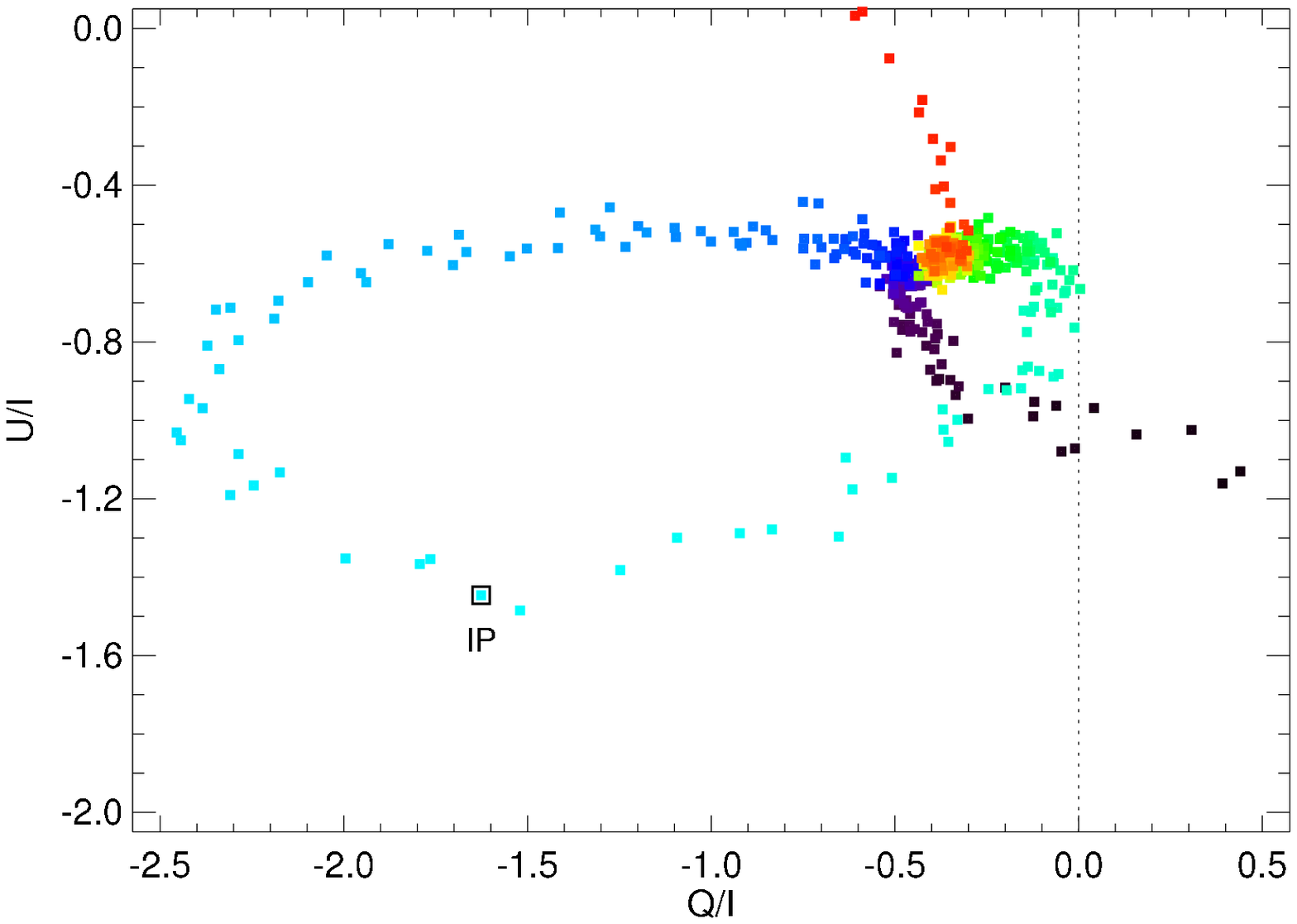}
\includegraphics[scale= 0.5]{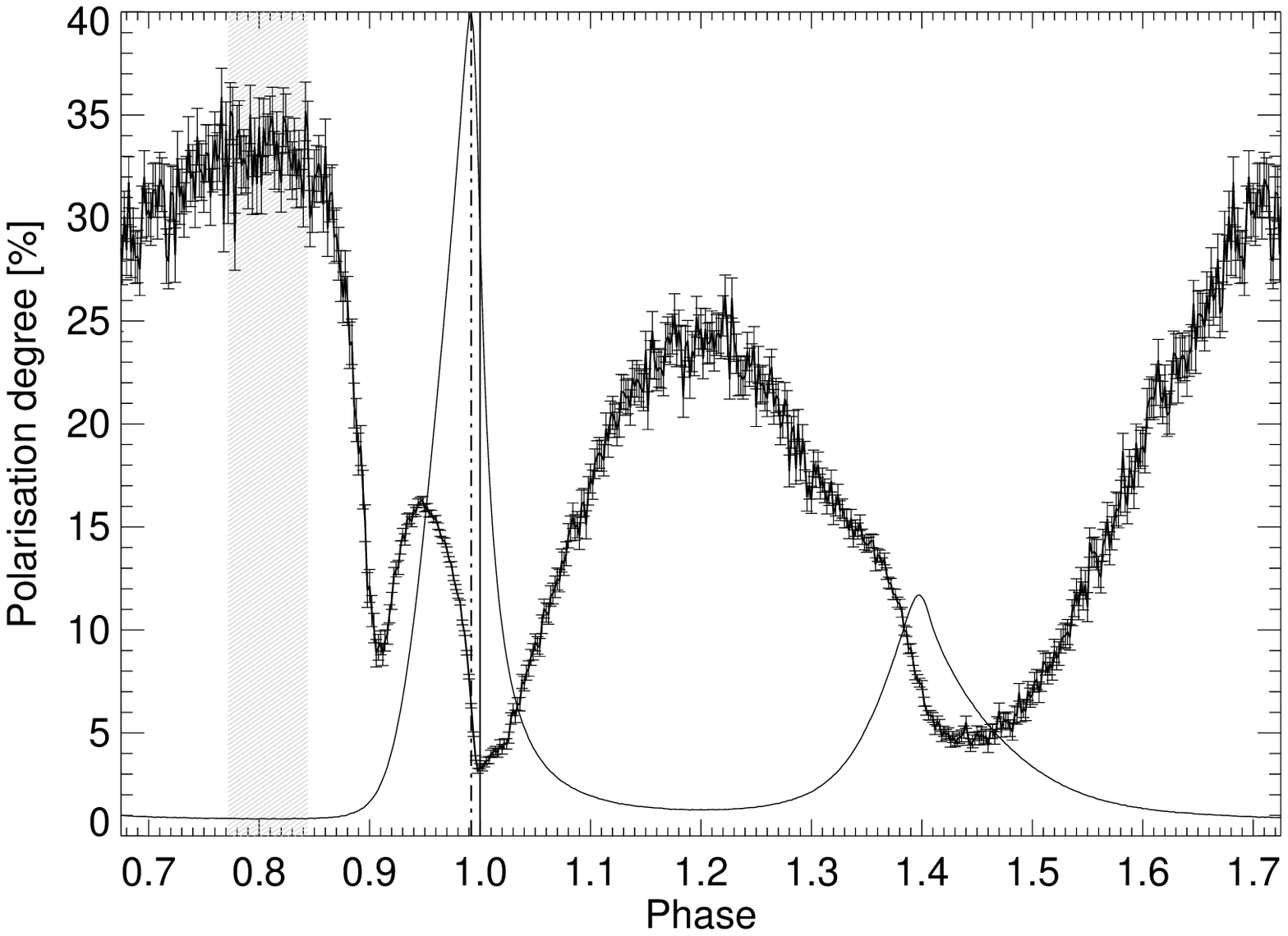}
\caption{\emph{Left column:}
Stokes parameters $Q$, $U$ as  a vector diagram. Colours refer to the
pulse phases as  indicated  in the input box of upper panel. The  scale is such
that $I=  100$ at  the maximum  light. Phases of  the MP  and IP  maxima are
indicated with  black open squares. Points belonging to  the MP and IP  follow
in a counter-clockwise     direction    on     outer    and     inner  ellipses,
respectively. Bottom panel shows a zoomed region around the IP phases. The dashed
lines indicate the origin of the scaled Q,U coordinates. Any error in background
polarisation would correspond to a shift in the origin in this plot. The lower
plot shows an zoomed region of the upper panel.
\emph{Right column:}
The  Crab pulsar position angle as a function of rotational phase (upper panel).
Changes of $\theta$  are aligned with the MP maximum of the optical light
(vertical dot-dashed line), but  also with the zero  phase, i.e.
with the radio phase of  the MP (vertical solid line, see Fig. \ref{Fig:zoom}
for details). The  Crab pulsar polarisation degree as a function of rotational
phase (bottom panel). Minimum of $p$ is for  the radio phase (vertical solid line),
and not for the maximum phase of the optical MP (vertical dot-dashed line, see
Fig. \ref{Fig:zoom} for details). For both panels a bit more than one rotation
is shown for clarity. The pulse profile (solid  line) and  DC phase range
(dashed region) are indicated also. The DC component in the phase
range 0.78--0.84 has P.A. $\sim 119\degr$, and P.D. $\sim 33\%$.
One sigma errors are show.}
\label{Fig:pa_pd}
\end{figure*}

\begin{figure*}
\includegraphics[scale= 0.63, angle= 90]{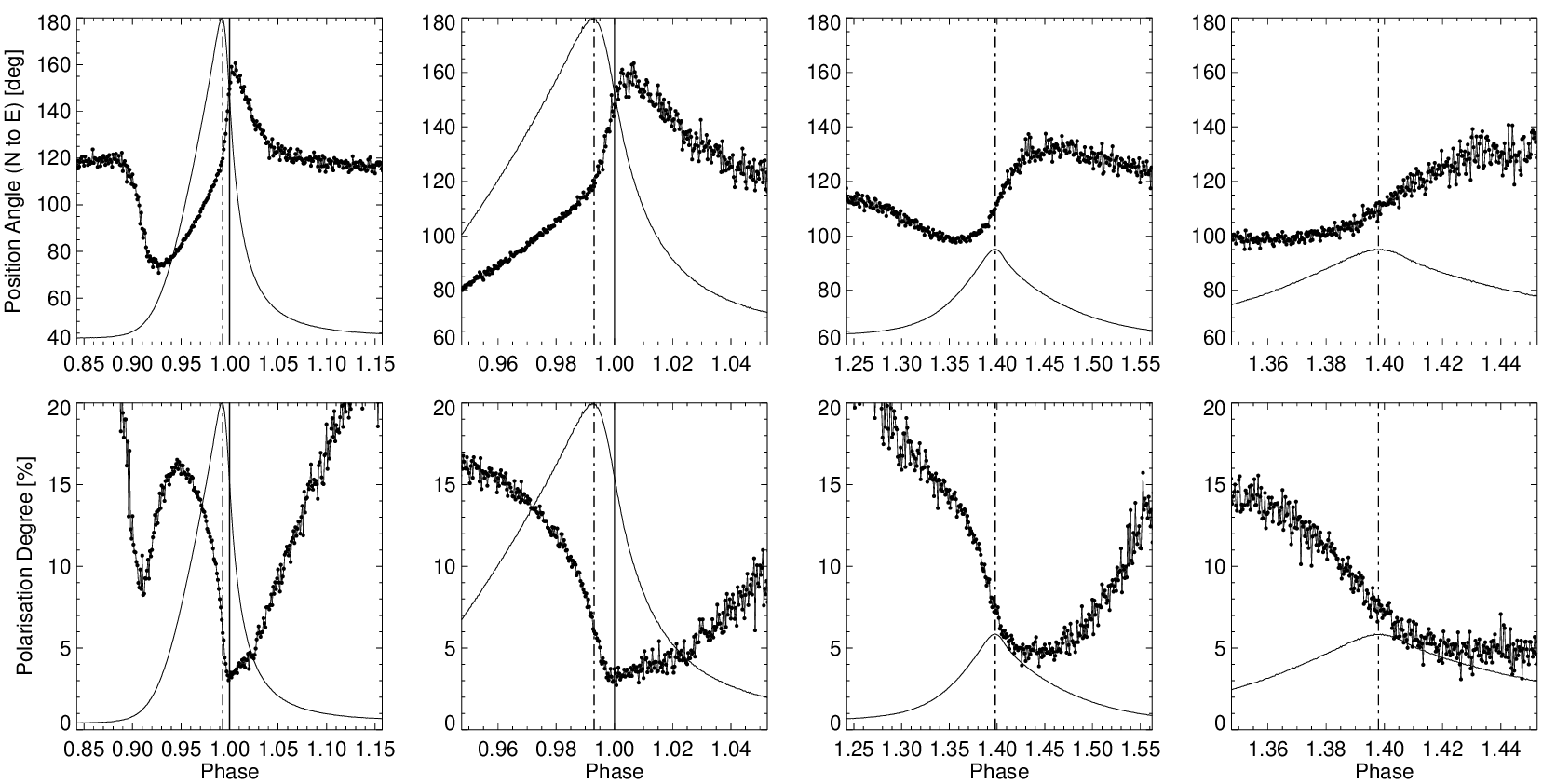}
\caption{Polarization characteristics of the Crab pulsar - P.A. (top row) and
  P.D. (bottom row) - as a function of rotational phase. Data binning is 1000
  and 3000 bins per cycle (corresponding to $33.5~\rm\mu s$ and $11~\rm\mu s$
  per bin) for odd and even columns, respectively. The first two columns are
  zoomed around the MP phase (using different scales), whereas the next two
  are zoomed around the IP phase. Dot-dashed vertical line indicates the
  optical MP maximum phase (=0.993), as well as the IP maximum phase (=
  1.398).  Solid vertical line indicates the radio peak phase (=1.0). For
  clarity the optical light curve of the Crab pulsar is over plotted (solid
  line).}
\label{Fig:zoom}
\end{figure*}

The Crab pulsar is detected at all phases of rotation, i.e. also in the
so-called `off-pulse' phase with an intensity of about 2\% compared to the
maximum intensity of the main pulse. As mentioned before, the measured level
of the DC differs from author to author.  For example very early measurements
performed by \citet{Peterson1978} gives the `unpulsed background' from the
Crab pulsar on the level of 3.6\% of the main peak intensity. Much lower
values were obtained by \citet{Jones1981} and later by \citet{Smith1988}:
0.6\% and 1.2\%, respectively.  In addition \citet{Percival1993} claims that
the `off-pulsed' flux has an intensity less than 0.9\% of the peak flux
(visible and UV data from $HST$, $2\sigma$ upper limit), whereas a fractional
flux derived by \citet{Golden2000off} from photometric analysis gives $\sim
1\%$.

For the  background (nebula and sky)  subtraction we took  the averaged
Stokes parameters  ($I_{DC}, {Q_{DC}}, U_{DC}$) recorded  in  the
ring  fibres  over  the `off-pulse'
phase. The pulsar Stokes parameters ($I$, $Q$, $U$; Fig.~\ref{Fig:stokes_norm})
as a function of its rotational phase are derived after subtracting
from the central channel measurements the steady nebular component.
The colour coded Stokes parameters   $Q$,   $U$   as   a  vector diagram
are   shown in the Fig.~\ref{Fig:pa_pd}. Colours refer to  the pulse phases
as indicated in the input box and the scale is such that $I= 100$ at the
maximum light, i.e maximum of the MP. The MP and IP maxima  are indicated
with black open squares. Points belonging to the MP phases follow  an  outer
ellipse (upper  panel),  whereas  these belonging to  the IP  an inner  one
(bottom panel).  In both  cases the direction of increasing pulsar phase
is counter-clockwise. Noteworthy
is  that already from  the Stokes parameters one can see that there
is sudden change in the pattern near the radio phase, i.e. where the red
points change to the black ones. As the next step of the data analysis
the polarisation characteristics, the position angle and the degree of
polarisation  of the E-vector, are calculated from the Stokes parameters
(Q.~\ref{Eq:pdpa}). Results plotted with a different time resolution
are presented in Fig.~\ref{Fig:pa_pd} and Fig.~\ref{Fig:zoom}.

Optical emission from the Crab pulsar is highly polarised, especially  in the
bridge and the `off-pulse' phases. The position angle and the degree of linear
polarisation as functions of rotation phase show well determined properties:
\begin{itemize}
\item[a)] The polarisation characteristics of both, the MP and the
IP components, are quite similar.
\item[b)] The polarisation degree reaches a minimum at phase close
to the {\em radio} main peak; the minimum  is {\em not}  aligned with
the optical peak.
\item[c)] There is a well defined bump in the polarisation degree
on the rising flank of the MP.
\item[d)] There is an indication of such a bump also for the IP
(especially after DC subtraction).
\item[e)] The position angle swings through a large angle in both
peaks: after subtraction of the DC component the angle swing is
130$\degr$ and 100$\degr$ for MP and IP, respectively.
\item[f)] The position angle at the bridge and `off-pulse' phases is
constant.
\item[g)] The position angle slope changes dramatically
at phases 0.993 (MP maximum) and 1.0 (radio peak).
\item[h)] The trailing wing of the MP (phase range 1.0-1.03)
shows a linearly increasing degree of polarisation. This feature
turns into a bump shape after DC subtraction. There is a slight
indication of the same behaviour for the trailing wing of the IP.
\end{itemize}

\subsection{Polarisation   characteristics  of  the   Crab  pulsar   after  DC
  subtraction}

\begin{figure*}
\includegraphics[scale= 0.63, angle= 90]{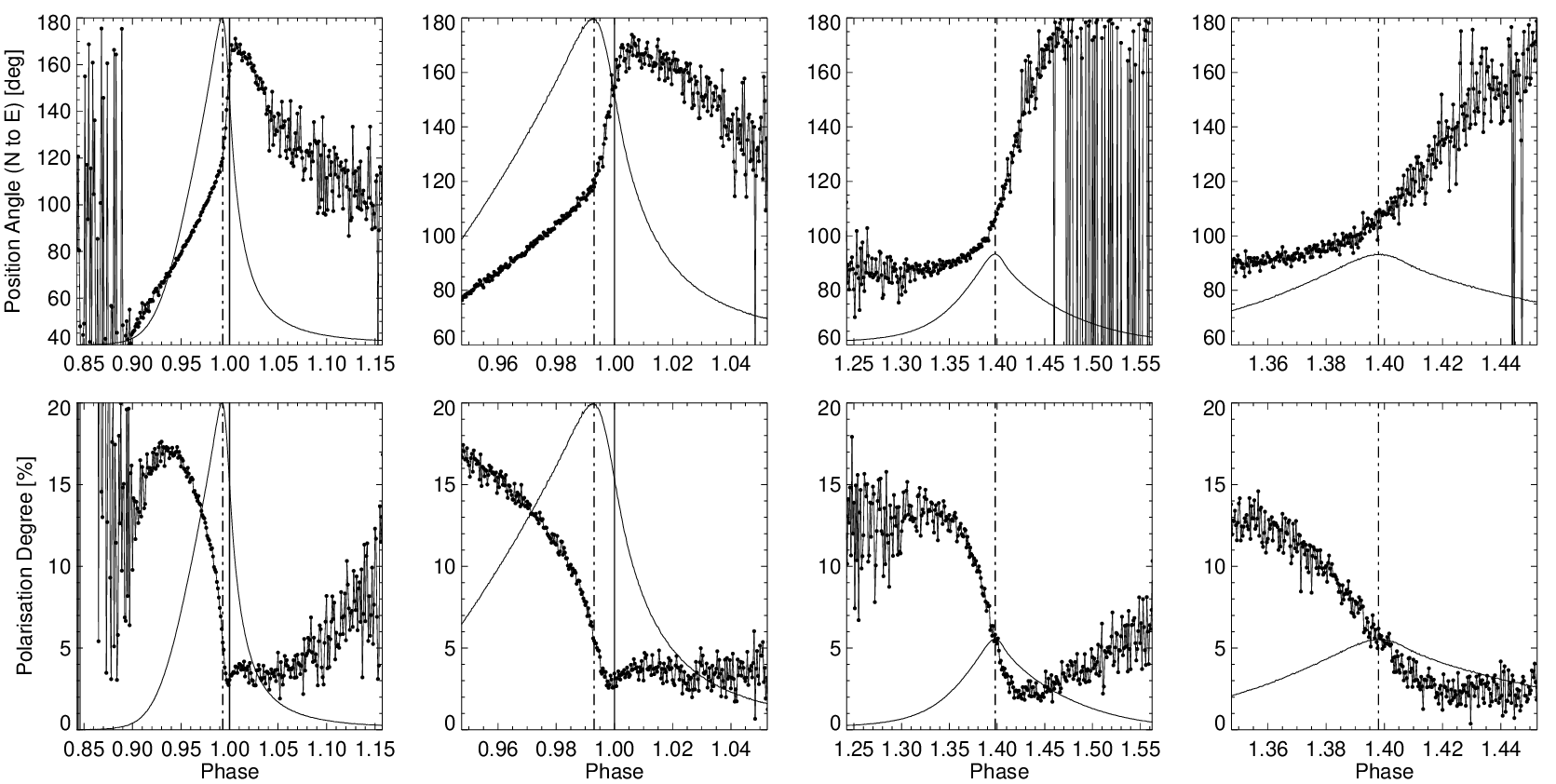}
\caption{The same as in Fig.~\ref{Fig:zoom} but after DC subtraction.}
\label{Fig:zoom_dc}
\end{figure*}

\begin{figure}
\includegraphics[scale= 0.5]{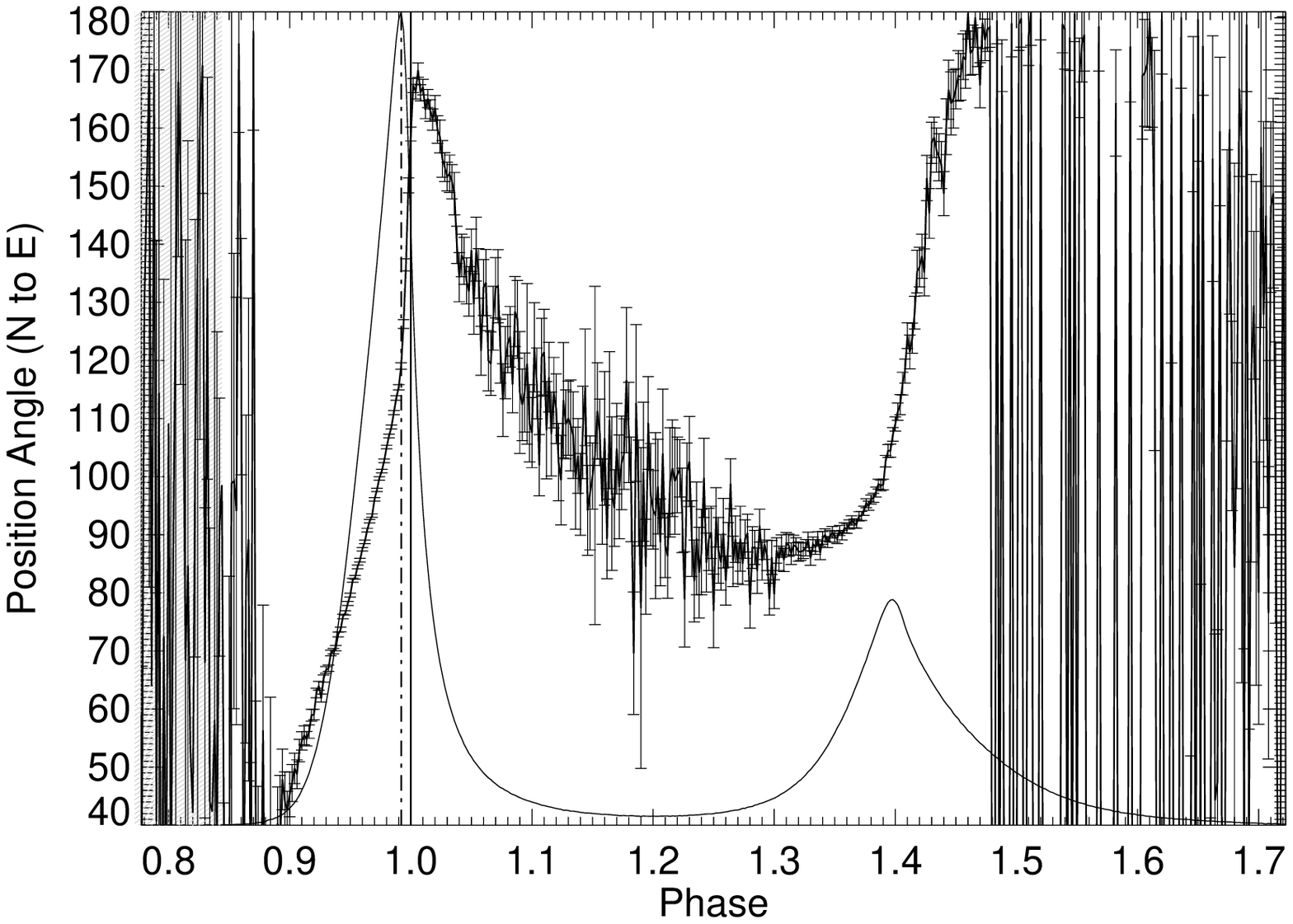}
\includegraphics[scale= 0.5]{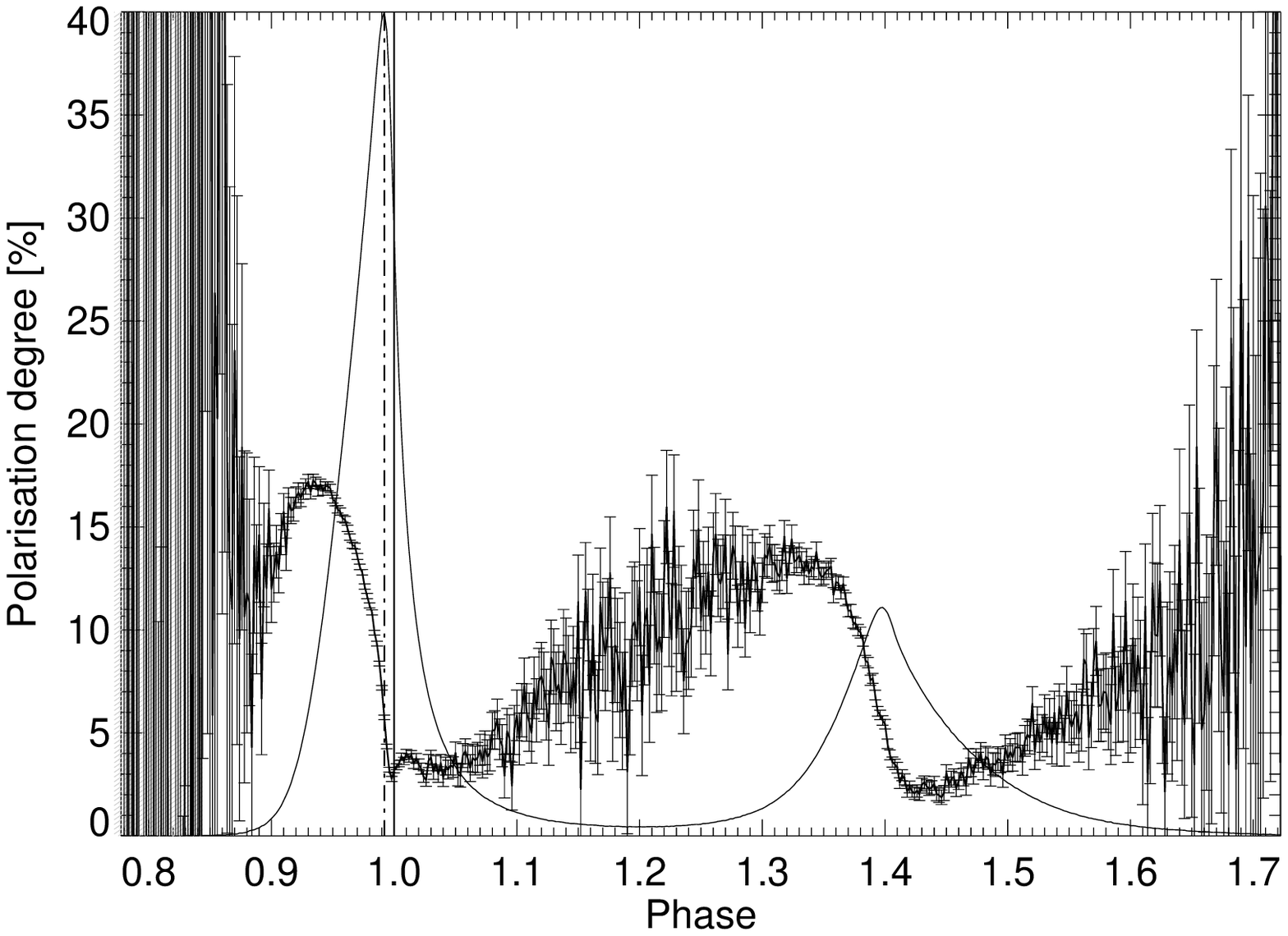}
\caption{The same as Fig. \ref{Fig:pa_pd} but after DC subtraction.}
\label{Fig:pa_pd_dc}
\end{figure}

The apparent constancy of the position angle within the phase range 0.78--0.84
(Fig. \ref{Fig:pa_pd}, right upper panel)  may suggest that the optical emission
from the  Crab pulsar consists  of two components  - pulsed and  unpulsed. The
pulsed  component is  characterised by  a highly  variable position  angle and
polarisation  degree. The  unpulsed (i.e.  DC) component  is  characterised by
constant intensity  on the  level of  2\% of the  main pulse  intensity, fixed
$\theta  \sim  119\degr$,  and  a  degree  of polarisation  on  the  level  of
33\%.  The source of emission of the  DC is unknown. There  are various ideas
from  where in  the magnetosphere  or  the  nebula this  component might
arise. It  is also  possible that the  inner knot (located  only $0\farcs65$
apart  from  the pulsar  and  being  a  persistent feature  throughout  the
sequences  of  the $HST$  images,  \cite{Hester2002}) contributes to the
`off-pulse' emission, although \citet{Golden2000off} claim that the Crab
image during the off-phase is compatible with an unresolved point source.
Assuming that the unpulsed component is  present at all
phase  angles and has constant  polarisation we  obtained  the polarisation
characteristics of  the `pulsed component' separately by  subtracting the
respective Stokes parameters $I_{DC}$, $Q_{DC}$, $U_{DC}$. Results are presented
in the Fig.~\ref{Fig:zoom_dc} and \ref{Fig:pa_pd_dc}. After subtracting the
`unpulsed component'
the position angle and polarisation degree in the phases where the intensity
is very low are not well defined. The values of $\theta$ and $p$
become very noisy because the Stokes parameters go basically
to zero for these rotational phases.

The polarisation  degree of the  `off-pulse' component obtained in  this work
differs    from     the     values     presented    by     other     authors:
e.g. \citet{Jones1981}: $\sim 70\%$; \citet{Smith1988}: $\sim  47\% \pm 10\% $.
This    might   be   caused    by   two reasons. Firstly, different groups treat
background subtraction in different ways. Additionally,  all mentioned
observations were  taken during different epochs  and  with different
instrumentation.  Secondly,  the observed  variation might  be  caused  by
the intrinsic  mechanism  of  the  pulsar  and/or  nebula radiation.
It should also be noted that the predicted decrease of the optical luminosity
is $\sim 1\%$ each two years \citep{Pacini1971}, therefore one  would  expect
a reduction of $0.13$~mag over the 22 years between \citet{Jones1981} and our
observations.

\subsection{Alignment between optical and radio wavelengths}

Precise timing of pulsar  light curves throughout the
electromagnetic spectrum can  be used  to constrain  theories of
the spatial distribution of various emission regions  and their
specific  propagation delays. In the  radio regime this  concept
has  been  applied  in the  so called frequency  mapping analysis.
From recent observations at different energies,  it became clear
that the Crab pulsar emission maxima of MP and IP are not aligned
in phase  at different wavelengths, from radio to the $\gamma$
energy range. Comparing  the visible and UV light curves obtained
from HST, \citet{Percival1993} were  among  the  first  to  show
that the  phase separation between the two peaks, ipso facto the
phases of the pulse maxima, change with energy. Since then many
authors have measured this effect in an attempt to understand  its
relation  to and   impact  on the  emission mechanism. However,
the techniques used by different authors to  measure the phase
separation (i.e. the phases  of the peak maxima) have varied.
Therefore,  this might cause method-dependent biases.
\citet{EikenberryAndFazio1997} showed that the peak-to-peak
separation appears to  be a more or less smooth function of
energy from infrared  to $\gamma$-ray  energies. The separation
decreases from $0.4087\pm0.0003$ to $0.398\pm0.003$ with energy
over the range from 0.5~eV to 1~MeV, respectively.  There  is
some evidence of  a turnover or a break in this trend at energies
of 0.7~eV ($H$ band  pass filter). No default method exists for
determining the position of the peak of the profile. For our
purpose and using  our high statistics light curve it was enough
to determine the peak  phases just by looking for  the maximum
intensity. This  gives us the phase  values  for the   MP:
$0.993\pm0.001$ and  the IP:  $0.398\pm0.001$ .  Our peak-to-peak
separation is  on the level of $0.4050\pm0.0014$. It  is in very
good agreement with the values obtained by
\citet{EikenberryAndFazio1997} for the visual  pass band, i.e.
$0.4057\pm0.0003$, and  from a previous OPTIMA observation 
of $0.4060\pm0.0003$ (\citet{Straubmeier01a}).

\begin{figure*}
\includegraphics[scale= 0.75, angle= 90]{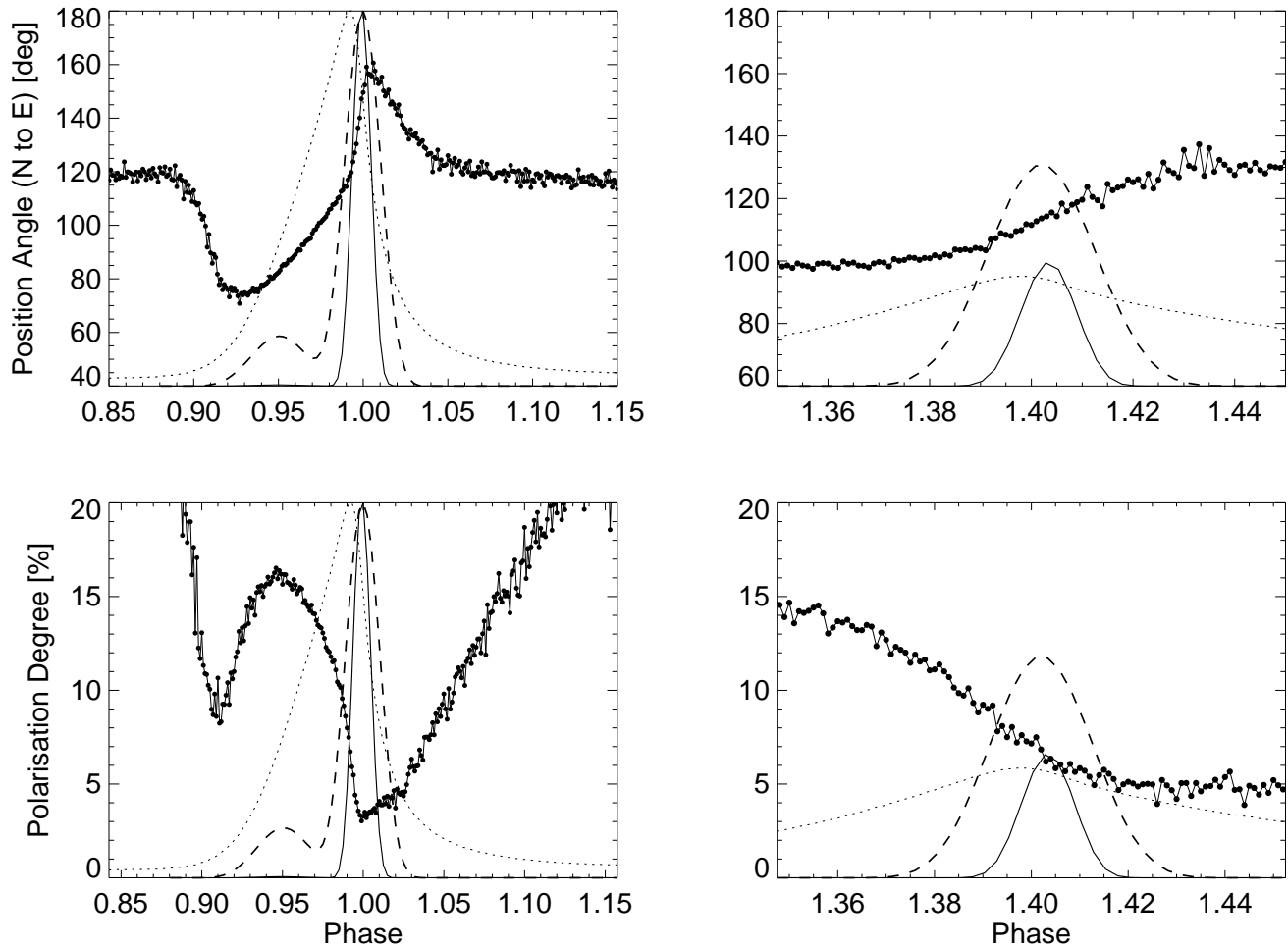}
\caption{Optical polarisation characteristics of the Crab pulsar (P.A. - top
row, P.D. - bottom row) compared with the pulsar radio profiles. Radio profiles
obtained at radio frequencies of 1400 MHz and 610 MHz are shown as a solid line
and dashed line, respectively. Left column shows zoom around the main pulse phase,
whereas right column around the inter pulse phase. Points indicate the optical
polarisation measurements, while the dotted line shows the optical intensity profile.}
\label{Fig:radio_and_optical}
\end{figure*}

Simultaneously with the optical observations we performed radio
observations at Jodrell Bank observatory, which allows us to
compare these wavelength ranges. Pulsed radio emission at 610 and
1400 MHz is detectable only around the peaks of emission due to
the strong background from the nebula.
Fig.~\ref{Fig:radio_and_optical} shows the radio light curves. At
610 MHz the MP is preceded by the so-called precursor peak, which
is mostly visible at lower frequencies and is not detected at 1400
MHz. The nature of this radio precursor is unclear - some
researchers proposed that it is the proper emission from the polar
gap of the pulsar while the main peaks are generated higher up in
the magnetosphere \citep[e.g.][]{Rankin1990}. 
Recently, Petrova (\citeyear{Petrova2008}) showed
that different components in the Crab pulse profile may be induced
by the scattering from different harmonics of the particle gyrofrequency
that takes place at different magnetospheric altitudes. This and
the rotational effect give rise to the components. In this model,
the low frequency component is formed by the scattering from the
first harmonic of the gyrofrequency into the state below the resonance,
whereas the precursor by the scattering between the states below the
resonance. This model is well supported by the radio polarization data.

As can be seen in Fig.~\ref{Fig:radio_and_optical} the optical
maxima of the MP are leading the radio pulse. A similar lead has
also been found at X- and $\gamma$-ray energies. The reported
values of the time lag are as follow: $344\pm40~\rm{\mu s}$
\citep[RXTE data]{Rots2004}, $280\pm40~\rm{\mu s}$ \citep[INTEGRAL
data]{Kuiper2003}, and $241\pm29~\rm{\mu s}$ \citep[EGRET
data]{Kuiper2003}. The uncertainty of the latter value does not
include the EGRET absolute timing uncertainty  of better  than
$100~\rm{\mu  s}$. At optical wavelengths  the situation  is quite
different and does not  provide  such   a coherent picture. The
quantitative amount of the lag between optical and radio peaks,
and whether it exists at all and at all times, is controversial
when we compare results obtained in different investigations.
Several authors reported the optical peak leading the radio peak
by a time shift of $140\pm78~\rm{\mu s}$ \citep{Sanwal1999},
$100\pm20~\rm{\mu s}$ \citep{Shearer2003}, and $273\pm100~\rm{\mu
s}$ recently found  by \cite {Oosterbroek2006}. On the other hand
\citet{Golden20002d} reported that the  optical pulse  trails  the
radio  pulse  by about  $80\pm60~\rm{\mu s}$. Additionally,
\citet{Romani2001} concluded that  both, radio and optical, peaks
are  coincident to better than $30~\rm{\mu  s}$, but this analysis
did not take  into account the uncertainty of  the radio ephemeris
being on the level  of $150~\rm{\mu s}$ in the  error
calculations.

From our measurements we conclude that the optical main peak is
leading the radio peak by a time shift  of  $231\pm68~\rm{\mu s}$.
The uncertainty in  this  value  is composed from an uncertainty
of $33~\rm{\mu s}$ in the determination of the optical peak of the
MP and from $60~\rm{\mu s}$ in the radio ephemeris. Our value of
the optical phase difference between  the  MP and IP of
$0.4050\pm0.0014$ is consistent with the latest  optical
measurements carried out by \citet{Oosterbroek2006}, who obtained
$0.4054\pm0.0004$. Both values, as  it  has already  been  shown
by \citet{EikenberryAndFazio1997}, are  not consistent  with  the
X-ray results,  e.g.  obtained  from   RXTE  data  by
\citet{Rots2004} of $0.4001\pm0.0002$. This implies  that the
details  of the pulse  profile in  X-rays and  in the optical
domain are  different.  In  a simple geometrical  model  (ignoring
relativistic  effects)  a  time  shift of $\sim 231~\rm{\mu s}$
indicates that possibly the optical radiation is formed $~\sim 70
~\rm{km}$  higher  in  the  magnetosphere than  the radio
emission. The difference in  phase of 0.007 could  also be
interpreted as  an angle  between the radio and optical beam of
$\sim 2.5 \degr$ (neglecting aberration and magnetic sweep back).

Another striking correlation between the radio intensity profile
and the optical polarisation can be seen in
Fig.~\ref{Fig:radio_and_optical}: the radio precursor seems to be
perfectly aligned with the bump in the degree of optical
polarisation. During this phase of the leading wing of the optical
MP the position angle change is also characterised by a nearly
linear swing. At the present stage of modelling the coherent and
incoherent emissions from a pulsar magnetosphere, where both
processes are generally treated independent of each other and
mutual interactions have not been investigated in depth, it is
premature to speculate on the origin of these observational
results.

\begin{figure*}
\includegraphics[scale= 0.75, angle= 90]{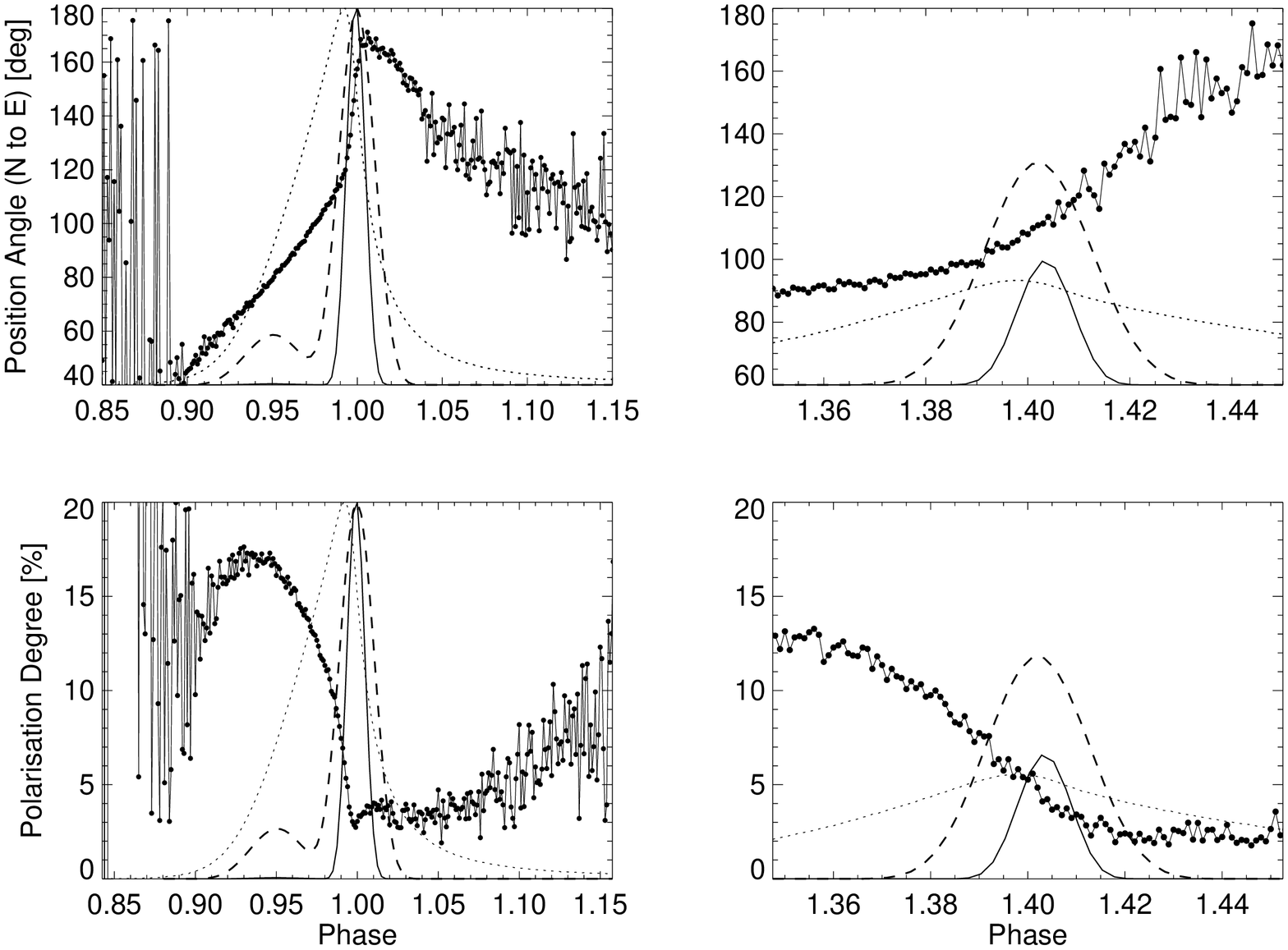}
\caption{The same as Fig.~\ref{Fig:radio_and_optical} but after DC subtraction.}
\label{Fig:radio_and_optical_dc}
\end{figure*}

\section{Summary and discussion}
\begin{table}
\caption{Polarisation degree (P.D.) and position angle (P.A.) at three selected Crab
pulsar phases, i.e. the phases of both optical maxima, IP (0.398) and
MP (0.993), as well as, the radio phase (0.000). Values before and
after DC component subtraction are shown.}
\label{Tab:phase_res}
\centering
\begin{tabular}{l c c c c }
\hline\hline
 & & & \multicolumn{2}{c}{After DC Subtraction} \\
Phase & P.D. [\%] & P.A. [$\degr$] & P.D. [\%] & P.A. [$\degr$] \\
\hline
0.398 &  $7.6\pm0.3$ & $109.9\pm1.1$ & $5.8\pm0.3$ & $106.0\pm1.6$  \\
0.993 &  $5.9\pm0.2$ & $119.8\pm0.8$ & $5.3\pm0.2$ & $119.9\pm0.8$  \\ 
0.000 &  $3.3\pm0.2$ & $149.6\pm1.6$ & $3.0\pm0.2$ & $157.5\pm1.7$  \\
\hline
\hline
\end{tabular}
\end{table}

\begin{figure*}
\includegraphics[scale=1.0]{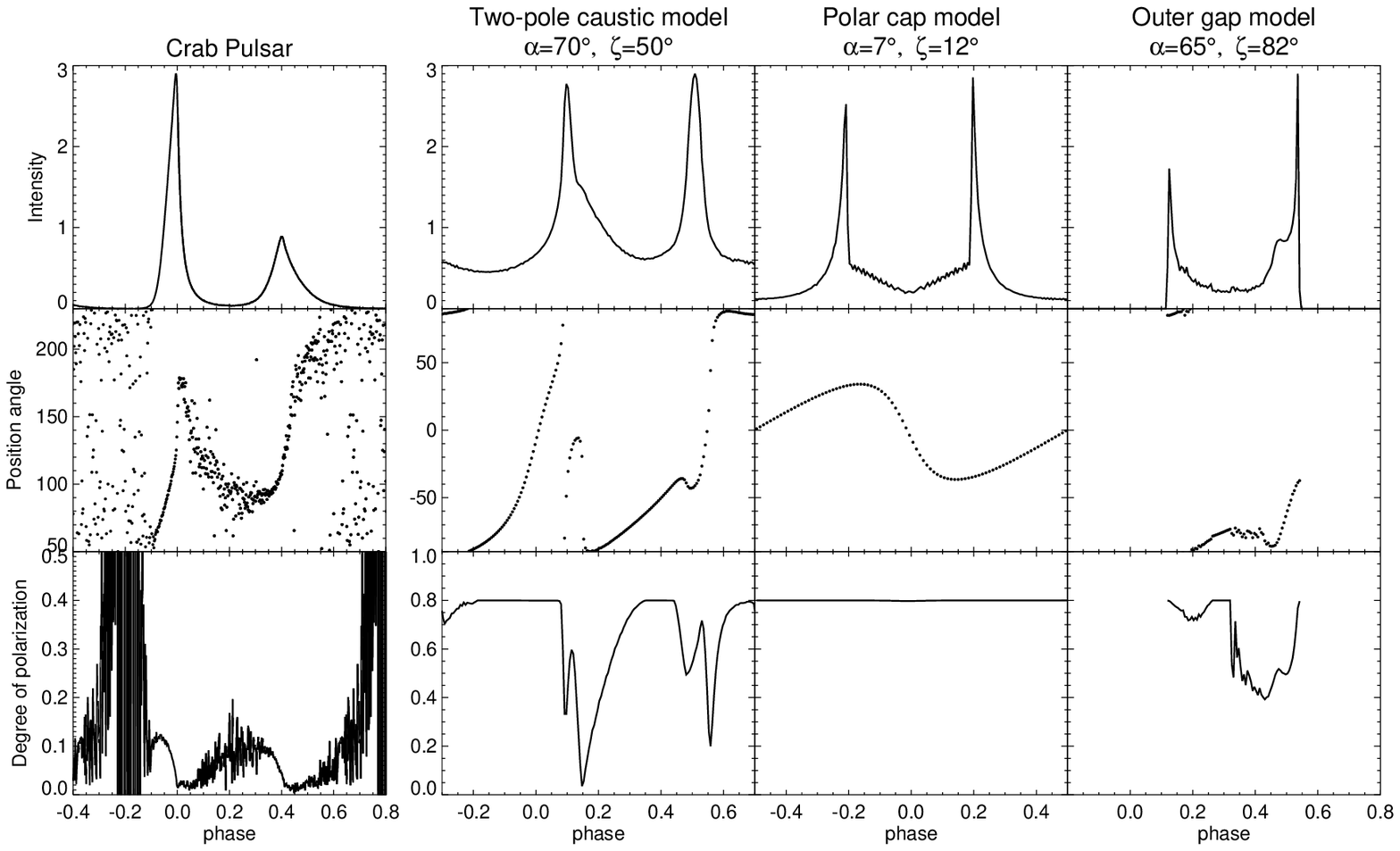}
\caption{The optical light curve, the position angle and the degree
of polarisation calculated with the following models of high energy radiation
from pulsars, from left to right: the polar cap model, the two-pole caustic model,
and the outer gap model \citep{Dyks2004IAUS}. In the first column, the data from
Calar Alto observations \citep{Kanbach2005} are shown for comparison. Courtesy J.
Dyks.}
\label{Fig:model_sydney}
\end{figure*}

The Crab pulsar emits highly anisotropic radiation which spans a wide
range of wavelengths, from  radio to  extreme $\gamma$-rays. Knowledge  of
polarisation characteristics of this radiation is of fundamental
importance in our attempts to  find the  mechanisms  responsible
for Crab's magnetospheric activity.  Good quality  X-ray  and $\gamma$-ray
polarimetry with  satellite  observatories  is expected to be available
for pulsar studies (among other types of objects) in the near future.
Present-day state of instrumentation allows to carry out optical
polarimetry of this object with unprecedented quality. Our project to
study the Crab pulsar with OPTIMA at NOT  is, to the best of our knowledge,
the most recent  and most complete one. The observations with
$11~\rm{\mu s}$ time resolution are an order of magnitude better than the
previous best observations. We have completely  resolved the polarisation
characteristics of both peaks of the Crab pulsar, MP and IP, in the optical
pass bands (see Fig.~\ref{Fig:zoom}). Moreover, we  were able  to better
characterise  the polarised  emission between the  peaks, i.e. the bridge
as well as the DC (`off-pulse') region.  We find that the MP of  the  Crab
pulsar  arrives $231\pm68~\rm{\mu s}$ before the peak of the radio pulse.

The phase averaged  polarisation degree of the Crab pulsar amounts to 9.8\%
with a  position angle of $109.5\degr$.  After  the  DC  subtraction  it is
$5.4\%$  and  $96.4\degr$, respectively. Minimum  polarisation degree
occurs  at the phase of  0.999, very  close to  the radio pulsar  phase, where
$p = 3.3\%\pm0.2\%, \theta  = 149.6 \degr\pm1.6\degr$ before the  DC subtraction,
and $p = 3.0\%\pm0.2\%, \theta = 157.5\degr\pm1.7\degr$ after DC subtraction.
During the IP the minimum value of $p$ is on the level of $5\%$, before DC
subtraction. Similar to the MP case, it is also shifted  (with respect to the
phase of the  optical maximum) to the leading wing  of the pulse. The value  of
$p$ changes after  DC subtraction to $\sim 2\%$.  This means that the situation
is inverted. Before the DC subtraction the  minimum of polarisation degree is 
reached during the main pulse, whereas after this subtraction it is  observed
during the  inter pulse.

The Crab is the pre-eminent example of a high-energy source with polarised
optical emission - if the pulsar is not phase resolved, the polarisations
still averages out to about 10\%. Other optical pulsars have been measured to
have an average polarisation also up to 10\%. Polarization can be generally
expected if the radiating particles are confined, e.g. by magnetic forces,
to anisotropic distributions. Detection of polarised objects in the field
of unidentified $\gamma$-ray sources could be therefore a valuable tracer
to aid identification.

Our   results   agree   generally   well   with   previous measurements
\citep[e.g.][]{Smith1988, Kanbach2005}, but they show details with much better
definition and statistics. The behaviour of $\theta$  as a function of phase
observed for  the Crab pulsar at optical wavelengths (Fig.~\ref{Fig:pa_pd})
differ from those observed  at radio wavelengths \citep[e.g.][]{Moffett1999,
Karastergiou2004, Slowikowska2005radio}. Two  factors may be responsible  for
this difference: different propagation effects, and different intrinsic emission
mechanisms. In particular, the  former factor plays an essential  role in such
high-energy emission models  like the outer  gap model or  the two-pole caustic
model: in both models  high-energy emission comes from  a very wide  range of
altitudes, contrary  to  radio  emission  which  originates  within  a  narrow
range  of altitudes. For  comparison the  light curves and  polarisation
characteristics obtained  within the framework  of three  high energy
magnetospheric emission models of pulsars,  i.e. the polar cap model, the
two-pole caustic model, and the outer gap model are shown in
Fig.~\ref{Fig:model_sydney} \citep{Dyks2004, Dyks2004IAUS}.

The  two-pole  caustic model  \citep{Dyks2003}  predicts  fast  swings of  the
position angle and minima in the polarisation degree,  similar to what is
observed. Polar cap model  and  outer  gap model  are  not  able  to reproduce
the  observational polarisation characteristics  of the Crab pulsar.  Another
model, placing the origin of the  pulsed optical emission from the Crab in  a
striped pulsar wind zone has been proposed by \cite*{Petri2005}. This model
features also polarisation characteristics that bear a certain resemblance 
to  the observations.

Recently, \citet{Takata2007}  attempted to simultaneously  reproduce all known
high energy emission properties of the Crab, including its optical polarisation
characteristics within the framework of a modified outer gap model. The model
is  restricted   to  synchrotron  emission  due  to   secondary  and  tertiary
electron-positron pairs which are expected in different spatial locations of the
3D gap; internal polarisation  characteristics are  calculated with  particular care. 
Yet, the calculated polarisation   properties   for   optical light  hardly   reproduce
the   observed properties. However, similarly  like in  the case  of the  two-pole
caustic  model \citep{Dyks2004, Dyks2004IAUS}, the  polarisation characteristics
obtained  in \citet{Takata2007}  become more consistent with  the  Crab optical
data  after  the  DC  component is subtracted   (as mentioned in  the previous
section).  An important outcome of \citet{Takata2007}  is that it offers the energy
dependence of the  polarisation features, covering the energy range between
1~eV and 10~keV.

The models of pulsar magnetospheric activity are based on various (sometimes ad
hoc) assumptions and different boundary conditions. Those lead to the model
differences in the macro scale (spatial extent of accelerators and emitting
regions), as  well as in the  micro scale (specific  radiative processes). The
former  include polar gaps, slot gaps, caustic gaps, outer gaps and striped winds.
The latter  include e.g.  curvature radiation,  synchrotron radiation  and inverse
Compton  scattering. In  consequence, the models differ significantly  in the
resulting `observed' radiation properties: light curves, energy spectra, and - last
but not least - polarisation \citep{Dyks2004, Dyks2004IAUS, Petri2005, Romani1995,
Takata2007}.

Linear polarisation characteristics in the high energy domain (optical, X-rays
and  $\gamma$-rays)  is  considered  as  a  powerful  tool  which  may  lead  to a
breakthrough  in our understanding of pulsar emission mechanism. For this reason,
the optical high time resolved polarisation properties obtained for the Crab pulsar
have attracted particular attention due to their uniqueness. Some models, like the
two-pole caustic model \citep{Dyks2004,  Dyks2004IAUS},   the  outer  gap   model 
\citep{Romani1995, Takata2007} or the striped pulsar wind model \citep{Petri2005}
are able to reproduce (very roughly) some of these properties, e.g. (a) and (e).
However, a fully convincing explanation of the properties listed from (a) to (h)
is beyond the reach of all above-mentioned models.

\section{Acknowledgements}
AS  acknowledges  support  from   the  EU  grant  MTKD-CT-2006  039965.  These
observations  were performed with the Nordic Optical Telescope, operated on the
island of La Palma  jointly  by Denmark,  Finland, Iceland,  Norway,  and Sweden,
in the  Spanish Observatorio del   Roque  de   los   Muchachos   of  the
Instituto   de  Astrofisica   de Canarias. Special thanks  to Fritz  Schrey from
MPE ,  as well as to  the staff from NOT  for very great hosting and all the help,
especially to Thomas  Augusteijn and Ingvar Sv\"ardh. We acknowledge the excellent 
support provided by  the NOT team. We would like to thank Bronek Rudak, Jarek
Dyks,  Ma\l{}gosia Sobolewska, as well as Aldo Serenelli and Andreas Zezas 
for useful discussions.

\bibliographystyle{mn2e}
\bibliography{slowikowska}

\begin{thebibliography}{}

\bibitem[\protect\citeauthoryear{{Aharonian}, , {Akhperjanian} \& {HEGRA
  Collaboration}}{{Aharonian} et~al.}{2004}]{Aharonian2004}
{Aharonian} F.,   {Akhperjanian} A.,    {HEGRA Collaboration} 2004, \apj, 614,
  897

\bibitem[\protect\citeauthoryear{{Aliu}, {Anderhub} \& {MAGIC
  Collaboration}}{{Aliu} et~al.}{2008}]{Aliu2008}
{Aliu} E.,  {Anderhub} H.,    {MAGIC Collaboration} 2008, Science, 322, 1221

\bibitem[\protect\citeauthoryear{{Allen}}{{Allen}}{2007}]{Allen2007PhD}
{Allen} B.~T.,  2007, PhD thesis, University of California, Irvine

\bibitem[\protect\citeauthoryear{{Allen}, {Yodh} \& {the Milagro
  Collaboration}}{{Allen} et~al.}{2007}]{Allen2007}
{Allen} B.~T.,  {Yodh} G.~B.,    {the Milagro Collaboration} 2007, Journal of
  Physics Conference Series, 60, 321

\bibitem[\protect\citeauthoryear{{Bridle}}{{Bridle}}{1970}]{Bridle1970}
{Bridle} A.~H.,  1970, \nat, 225, 1035

\bibitem[\protect\citeauthoryear{{Chanan} \& {Helfand}}{{Chanan} \&
  {Helfand}}{1990}]{Chanan1990}
{Chanan} G.~A.,  {Helfand} D.~J.,  1990, \apj, 352, 167

\bibitem[\protect\citeauthoryear{{Cocke}, {Disney}, {Muncaster} \&
  {Gehrels}}{{Cocke} et~al.}{1970}]{Cocke1970}
{Cocke} W.~J.,  {Disney} M.~J.,  {Muncaster} G.~W.,    {Gehrels} T.,  1970,
  \nat, 227, 1327

\bibitem[\protect\citeauthoryear{{Cocke}, {Ferguson} \& {Muncaster}}{{Cocke}
  et~al.}{1973}]{Cocke1973}
{Cocke} W.~J.,  {Ferguson} D.~C.,    {Muncaster} G.~W.,  1973, \apj, 183, 987

\bibitem[\protect\citeauthoryear{{Dean}, {Clark}, {Stephen}, {McBride},
  {Bassani}, {Bazzano}, {Bird}, {Hill}, {Shaw} \& {Ubertini}}{{Dean}
  et~al.}{2008}]{Dean2008}
{Dean} A.~J.,  {Clark} D.~J.,  {Stephen} J.~B.,  {McBride} V.~A.,  {Bassani}
  L.,  {Bazzano} A.,  {Bird} A.~J.,  {Hill} A.~B.,  {Shaw} S.~E.,    {Ubertini}
  P.,  2008, Science, 321, 1183

\bibitem[\protect\citeauthoryear{{Dyks}, {Harding} \& {Rudak}}{{Dyks}
  et~al.}{2004a}]{Dyks2004}
{Dyks} J.,  {Harding} A.~K.,    {Rudak} B.,  2004a, \apj, 606, 1125

\bibitem[\protect\citeauthoryear{{Dyks}, {Harding} \& {Rudak}}{{Dyks}
  et~al.}{2004b}]{Dyks2004IAUS}
{Dyks} J.,  {Harding} A.~K.,    {Rudak} B.,  2004b, in {Camilo} F.,  {Gaensler}
  B.~M.,  eds, IAU Symp. 218: Young Neutron Stars and Their Environments
  {Two-pole Caustic Model for High-energy Radiation from Pulsars -
  Polarization}.
pp 373--374

\bibitem[\protect\citeauthoryear{{Dyks} \& {Rudak}}{{Dyks} \&
  {Rudak}}{2003}]{Dyks2003}
{Dyks} J.,  {Rudak} B.,  2003, \apj, 598, 1201

\bibitem[\protect\citeauthoryear{{Eikenberry} \& {Fazio}}{{Eikenberry} \&
  {Fazio}}{1997}]{EikenberryAndFazio1997}
{Eikenberry} S.~S.,  {Fazio} G.~G.,  1997, \apj, 476, 281

\bibitem[\protect\citeauthoryear{{Ferguson}}{{Ferguson}}{1973}]{Ferguson1973}
{Ferguson} D.~C.,  1973, \apj, 183, 977

\bibitem[\protect\citeauthoryear{{Ferguson}, {Cocke} \& {Gehrels}}{{Ferguson}
  et~al.}{1974}]{Ferguson1974}
{Ferguson} D.~C.,  {Cocke} W.~J.,    {Gehrels} T.,  1974, \apj, 190, 375

\bibitem[\protect\citeauthoryear{{Forot}, {Laurent}, {Grenier}, {Gouiff{\`e}s}
  \& {Lebrun}}{{Forot} et~al.}{2008}]{Forot2008}
{Forot} M.,  {Laurent} P.,  {Grenier} I.~A.,  {Gouiff{\`e}s} C.,    {Lebrun}
  F.,  2008, \apjl, 688, L29

\bibitem[\protect\citeauthoryear{{Golden}, {Shearer} \& {Beskin}}{{Golden}
  et~al.}{2000}]{Golden2000off}
{Golden} A.,  {Shearer} A.,    {Beskin} G.~M.,  2000, \apj, 535, 373

\bibitem[\protect\citeauthoryear{{Golden}, {Shearer}, {Redfern}, {Beskin},
  {Neizvestny}, {Neustroev}, {Plokhotnichenko} \& {Cullum}}{{Golden}
  et~al.}{2000}]{Golden20002d}
{Golden} A.,  {Shearer} A.,  {Redfern} R.~M.,  {Beskin} G.~M.,  {Neizvestny}
  S.~I.,  {Neustroev} V.~V.,  {Plokhotnichenko} V.~L.,    {Cullum} M.,  2000,
  \aap, 363, 617

\bibitem[\protect\citeauthoryear{{Gould} \& {Lyne}}{{Gould} \&
  {Lyne}}{1998}]{Gould1998}
{Gould} D.~M.,  {Lyne} A.~G.,  1998, \mnras, 301, 235

\bibitem[\protect\citeauthoryear{{Graham-Smith}, {Dolan}, {Boyd}, {Biggs},
  {Lyne} \& {Percival}}{{Graham-Smith} et~al.}{1996}]{Graham-Smith1996}
{Graham-Smith} F.,  {Dolan} J.~F.,  {Boyd} P.~T.,  {Biggs} J.~D.,  {Lyne}
  A.~G.,    {Percival} J.~W.,  1996, \mnras, 282, 1354

\bibitem[\protect\citeauthoryear{{Hester}, {Mori}, {Burrows}, {Gallagher},
  {Graham}, {Halverson}, {Kader}, {Michel} \& {Scowen}}{{Hester}
  et~al.}{2002}]{Hester2002}
{Hester} J.~J.,  {Mori} K.,  {Burrows} D.,  {Gallagher} J.~S.,  {Graham} J.~R.,
   {Halverson} M.,  {Kader} A.,  {Michel} F.~C.,    {Scowen} P.,  2002, \apjl,
  577, L49

\bibitem[\protect\citeauthoryear{{Hester}, {Scowen}, {Sankrit} \& et
  al.}{{Hester} et~al.}{1995}]{Hester1995}
{Hester} J.~J.,  {Scowen} P.~A.,  {Sankrit} R.,    et al. 1995, \apj, 448, 240

\bibitem[\protect\citeauthoryear{{Jones}, {Smith} \& {Wallace}}{{Jones}
  et~al.}{1981}]{Jones1981}
{Jones} D.~H.~P.,  {Smith} F.~G.,    {Wallace} P.~T.,  1981, \mnras, 196, 943

\bibitem[\protect\citeauthoryear{{Kanbach}, {Kellner}, {Schrey}, {Steinle},
  {Straubmeier} \& {Spruit}}{{Kanbach} et~al.}{2003}]{Kanbach2003}
{Kanbach} G.,  {Kellner} S.,  {Schrey} F.~Z.,  {Steinle} H.,  {Straubmeier} C.,
     {Spruit} H.~C.,  2003, in {Iye} M.,  {Moorwood} A.~F.~M.,  eds, SPIE Proc.
  Instrument Design and Performance for Optical/Infrared Ground-based
  Telescopes Vol.~4841, {Design and results of the fast timing
  photo-polarimeter OPTIMA}.
pp 82--93

\bibitem[\protect\citeauthoryear{{Kanbach}, {S\l{}owikowska}, {Kellner} \&
  {Steinle}}{{Kanbach} et~al.}{2005}]{Kanbach2005}
{Kanbach} G.,  {S\l{}owikowska} A.,  {Kellner} S.,    {Steinle} H.,  2005, in
  {Bulik} T.,  {Rudak} B.,   {Madejski} G.,  eds, AIP Conf. Proc. 801:
  Astrophysical Sources of High Energy Particles and Radiation {New optical
  polarization measurements of the Crab pulsar}.
pp 306--311

\bibitem[\protect\citeauthoryear{{Kanbach}, {Stefanescu}, {Duscha}, {Steinle},
  {Burwitz} \& {Schwope}}{{Kanbach} et~al.}{2008}]{Kanbach2008}
{Kanbach} G.,  {Stefanescu} A.,  {Duscha} S.,  {Steinle} H.,  {Burwitz} V.,
  {Schwope} A.,  2008, in {Phelan} D.,  {Ryan} O.,   {Shearer} A.,  eds, High
  Time Resolution Astrophysics: The Universe at Sub-Second Timescales Vol.~984
  of American Institute of Physics Conference Series, {High time resolution
  observations of Cataclysmic Variables with OPTIMA}.
pp 32--40

\bibitem[\protect\citeauthoryear{{Karastergiou}, {Jessner} \&
  {Wielebinski}}{{Karastergiou} et~al.}{2004}]{Karastergiou2004}
{Karastergiou} A.,  {Jessner} A.,    {Wielebinski} R.,  2004, in {Camilo} F.,
  {Gaensler} B.~M.,  eds, IAU Symp. 218: Young Neutron Stars and Their
  Environments {High-frequency Polarimetric Observations of the Crab Pulsar}.
pp 329--330

\bibitem[\protect\citeauthoryear{{Karastergiou} \& {Johnston}}{{Karastergiou}
  \& {Johnston}}{2006}]{Karastergiou2006}
{Karastergiou} A.,  {Johnston} S.,  2006, \mnras, 365, 353

\bibitem[\protect\citeauthoryear{Kellner}{Kellner}{2002}]{Kellner2002}
Kellner S.,  2002, Master's thesis, TU-M\"unchen

\bibitem[\protect\citeauthoryear{{Kern}, {Martin}, {Mazin} \& {Halpern}}{{Kern}
  et~al.}{2003}]{Kern2003}
{Kern} B.,  {Martin} C.,  {Mazin} B.,    {Halpern} J.~P.,  2003, \apj, 597,
  1049

\bibitem[\protect\citeauthoryear{{Kristian}, {Visvanathan}, {Westphal} \&
  {Snellen}}{{Kristian} et~al.}{1970}]{Kristian1970}
{Kristian} J.,  {Visvanathan} N.,  {Westphal} J.~A.,    {Snellen} G.~H.,  1970,
  \apj, 162, 475

\bibitem[\protect\citeauthoryear{{Kuiper}, {Hermsen}, {Walter} \&
  {Foschini}}{{Kuiper} et~al.}{2003}]{Kuiper2003}
{Kuiper} L.,  {Hermsen} W.,  {Walter} R.,    {Foschini} L.,  2003, \aap, 411,
  L31

\bibitem[\protect\citeauthoryear{{Lyne} \& {Graham-Smith}}{{Lyne} \&
  {Graham-Smith}}{2006}]{Lyne2006}
{Lyne} A.~G.,  {Graham-Smith} F.,  2006, {Pulsar astronomy}.
Pulsar astronomy, 3rd ed., by A.G.~Lyne and F.~Graham-Smith.~Cambridge
  astrophysics series.~Cambridge, UK: Cambridge University Press, 2006 ISBN
  0521839548.

\bibitem[\protect\citeauthoryear{{Lyne}, {Pritchard} \& {Graham-Smith}}{{Lyne}
  et~al.}{1993}]{Lyne1993}
{Lyne} A.~G.,  {Pritchard} R.~S.,    {Graham-Smith} F.,  1993, \mnras, 265,
  1003

\bibitem[\protect\citeauthoryear{{Manchester}, {Hobbs}, {Teoh} \&
  {Hobbs}}{{Manchester} et~al.}{2005}]{Manchester2005}
{Manchester} R.~N.,  {Hobbs} G.~B.,  {Teoh} A.,    {Hobbs} M.,  2005, \aj, 129,
  1993

\bibitem[\protect\citeauthoryear{{Mazzuca}, {Sparks} \& {Axon}}{{Mazzuca}
  et~al.}{1998}]{Mazzuca1998}
{Mazzuca} L.,  {Sparks} W.~B.,    {Axon} D.,  1998, Instrument Science Report
  NICMOS 98-017

\bibitem[\protect\citeauthoryear{{McLean}, {Aspin} \& {Reitsema}}{{McLean}
  et~al.}{1983}]{McLean1983}
{McLean} I.~S.,  {Aspin} C.,    {Reitsema} H.,  1983, \nat, 304, 243

\bibitem[\protect\citeauthoryear{{Middleditch}, {Pennypacker} \&
  {Burns}}{{Middleditch} et~al.}{1987}]{Middleditch1987}
{Middleditch} J.,  {Pennypacker} C.~R.,    {Burns} M.~S.,  1987, \apj, 315, 142

\bibitem[\protect\citeauthoryear{{Mignani}, {Bagnulo}, {Dyks}, {Lo Curto} \&
  {S\l{}owikowska}}{{Mignani} et~al.}{2007}]{Mignani2007}
{Mignani} R.~P.,  {Bagnulo} S.,  {Dyks} J.,  {Lo Curto} G.,    {S\l{}owikowska}
  A.,  2007, \aap, 467, 1157

\bibitem[\protect\citeauthoryear{{Moffett} \& {Hankins}}{{Moffett} \&
  {Hankins}}{1999}]{Moffett1999}
{Moffett} D.~A.,  {Hankins} T.~H.,  1999, \apj, 522, 1046

\bibitem[\protect\citeauthoryear{M\"uhlegger}{M\"uhlegger}{2006}]{Muehlegger20%
06}
M\"uhlegger M.,  2006, Master's thesis, TU-M\"unechen

\bibitem[\protect\citeauthoryear{{Ng} \& {Romani}}{{Ng} \&
  {Romani}}{2006}]{Ng2006}
{Ng} C.-Y.,  {Romani} R.~W.,  2006, \apj, 644, 445

\bibitem[\protect\citeauthoryear{{Oort} \& {Walraven}}{{Oort} \&
  {Walraven}}{1956}]{Oort1956}
{Oort} J.~H.,  {Walraven} T.,  1956, \bain, 12, 285

\bibitem[\protect\citeauthoryear{Oosterbroek, {de Bruijne}, Martin, Verhoeve,
  Perryman, Erd \& Schulz}{Oosterbroek et~al.}{2006}]{Oosterbroek2006}
Oosterbroek T.,  {de Bruijne} J.~H.~J.,  Martin D.,  Verhoeve P.,  Perryman
  M.~A.~C.,  Erd C.,    Schulz R.,  2006, astro-ph/0606146

\bibitem[\protect\citeauthoryear{{Pacini}}{{Pacini}}{1971}]{Pacini1971}
{Pacini} F.,  1971, \apjl, 163, L17

\bibitem[\protect\citeauthoryear{{Percival}, {Biggs}, {Dolan}, {Robinson},
  {Taylor}, {Bless}, {Elliot}, {Nelson}, {Ramseyer}, {van Citters} \&
  {Zhang}}{{Percival} et~al.}{1993}]{Percival1993}
{Percival} J.~W.,  {Biggs} J.~D.,  {Dolan} J.~F.,  {Robinson} E.~L.,  {Taylor}
  M.~J.,  {Bless} R.~C.,  {Elliot} J.~L.,  {Nelson} M.~J.,  {Ramseyer} T.~F.,
  {van Citters} G.~W.,    {Zhang} E.,  1993, \apj, 407, 276

\bibitem[\protect\citeauthoryear{{Peterson}, {Murdin}, {Wallace}, {Manchester},
  {Penny}, {Jorden}, {Hartley} \& {King}}{{Peterson}
  et~al.}{1978}]{Peterson1978}
{Peterson} B.~A.,  {Murdin} P.,  {Wallace} P.,  {Manchester} R.~N.,  {Penny}
  A.~J.,  {Jorden} A.,  {Hartley} K.~F.,    {King} D.,  1978, \nat, 276, 475

\bibitem[\protect\citeauthoryear{{P{\'e}tri} \& {Kirk}}{{P{\'e}tri} \&
  {Kirk}}{2005}]{Petri2005}
{P{\'e}tri} J.,  {Kirk} J.~G.,  2005, \apjl, 627, L37

\bibitem[\protect\citeauthoryear{{Petrova}}{{Petrova}}{2008}]{Petrova2008}
{Petrova} S.~A.,  2008, \mnras, 385, 2143

\bibitem[\protect\citeauthoryear{{Radhakrishnan} \& {Cooke}}{{Radhakrishnan} \&
  {Cooke}}{1969}]{Radhakrishnan1969}
{Radhakrishnan} V.,  {Cooke} D.~J.,  1969, \aplett, 3, 225

\bibitem[\protect\citeauthoryear{{Rankin}}{{Rankin}}{1990}]{Rankin1990}
{Rankin} J.~M.,  1990, \apj, 352, 247

\bibitem[\protect\citeauthoryear{{Romani}, {Miller}, {Cabrera}, {Nam} \&
  {Martinis}}{{Romani} et~al.}{2001}]{Romani2001}
{Romani} R.~W.,  {Miller} A.~J.,  {Cabrera} B.,  {Nam} S.~W.,    {Martinis}
  J.~M.,  2001, \apj, 563, 221

\bibitem[\protect\citeauthoryear{{Romani} \& {Yadigaroglu}}{{Romani} \&
  {Yadigaroglu}}{1995}]{Romani1995}
{Romani} R.~W.,  {Yadigaroglu} I.-A.,  1995, \apj, 438, 314

\bibitem[\protect\citeauthoryear{{Rots}, {Jahoda} \& {Lyne}}{{Rots}
  et~al.}{2004}]{Rots2004}
{Rots} A.~H.,  {Jahoda} K.,    {Lyne} A.~G.,  2004, \apjl, 605, L129

\bibitem[\protect\citeauthoryear{Sanwal}{Sanwal}{1999}]{Sanwal1999}
Sanwal D.,  1999, PhD thesis, The University of Texas at Austin

\bibitem[\protect\citeauthoryear{{Shearer}, {Stappers}, {O'Connor}, {Golden},
  {Strom}, {Redfern} \& {Ryan}}{{Shearer} et~al.}{2003}]{Shearer2003}
{Shearer} A.,  {Stappers} B.,  {O'Connor} P.,  {Golden} A.,  {Strom} R.,
  {Redfern} M.,    {Ryan} O.,  2003, Science, 301, 493

\bibitem[\protect\citeauthoryear{{Silver}, {Kestenbaum}, {Long}, {Novick},
  {Wolff} \& {Weisskopf}}{{Silver} et~al.}{1978}]{Silver1978}
{Silver} E.~H.,  {Kestenbaum} H.~L.,  {Long} K.~S.,  {Novick} R.,  {Wolff}
  R.~S.,    {Weisskopf} M.~C.,  1978, \apj, 225, 221

\bibitem[\protect\citeauthoryear{{S\l{}owikowska}, {Jessner}, {Klein} \&
  {Kanbach}}{{S\l{}owikowska} et~al.}{2005}]{Slowikowska2005radio}
{S\l{}owikowska} A.,  {Jessner} A.,  {Klein} B.,    {Kanbach} G.,  2005, in
  {Bulik} T.,  {Rudak} B.,   {Madejski} G.,  eds, AIP Conf. Proc. 801:
  Astrophysical Sources of High Energy Particles and Radiation {Polarization
  characteristics of the Crab pulsar's giant radio pulses at HFCs phases}.
pp 324--329

\bibitem[\protect\citeauthoryear{{S\l{}owikowska}, {Rudak} \&
  {Kanbach}}{{S\l{}owikowska} et~al.}{2008}]{Slowikowska2008}
{S\l{}owikowska} A.,  {Rudak} B.,    {Kanbach} G.,  2008, in {Bassa} C.,
  {Wang} Z.,  {Cumming} A.,   {Kaspi} V.~M.,  eds, 40 Years of Pulsars:
  Millisecond Pulsars, Magnetars and More Vol.~983 of American Institute of
  Physics Conference Series, {High energy polarization of pulsars-observations
  vs. models}.
pp 142--144

\bibitem[\protect\citeauthoryear{{Smith}, {Jones}, {Dick} \& {Pike}}{{Smith}
  et~al.}{1988}]{Smith1988}
{Smith} F.~G.,  {Jones} D.~H.~P.,  {Dick} J.~S.~B.,    {Pike} C.~D.,  1988,
  \mnras, 233, 305

\bibitem[\protect\citeauthoryear{{Sparks} \& {Axon}}{{Sparks} \&
  {Axon}}{1999}]{Sparks1999}
{Sparks} W.~B.,  {Axon} D.~J.,  1999, \pasp, 111, 1298

\bibitem[\protect\citeauthoryear{{Standish}
  Jr.}{{Standish}}{1982}]{Standish1982}
{Standish} Jr. E.~M.,  1982, \aap, 114, 297

\bibitem[\protect\citeauthoryear{Straubmeier}{Straubmeier}{2001}]{Straubmeier0%
1a}
Straubmeier C.,  2001, PhD thesis, TU-M\"unchen

\bibitem[\protect\citeauthoryear{{Takata}, {Chang} \& {Cheng}}{{Takata}
  et~al.}{2007}]{Takata2007}
{Takata} J.,  {Chang} H.-K.,    {Cheng} K.~S.,  2007, \apj, 656, 1044

\bibitem[\protect\citeauthoryear{{Turnshek}, {Bohlin}, {Williamson} II,
  {Lupie}, {Koornneef} \& {Morgan}}{{Turnshek} et~al.}{1990}]{Turnshek1990}
{Turnshek} D.~A.,  {Bohlin} R.~C.,  {Williamson} II R.~L.,  {Lupie} O.~L.,
  {Koornneef} J.,    {Morgan} D.~H.,  1990, \aj, 99, 1243

\bibitem[\protect\citeauthoryear{{Wagner} \& {Seifert}}{{Wagner} \&
  {Seifert}}{2000}]{Wagner2000}
{Wagner} S.~J.,  {Seifert} W.,  2000, in {Kramer} M.,  {Wex} N.,
  {Wielebinski} R.,  eds, IAU Colloq. 177: Pulsar Astronomy - 2000 and Beyond
  Vol.~202 of Astronomical Society of the Pacific Conference Series, {Optical
  Polarization Measurements of Pulsars}.
pp 315--+

\bibitem[\protect\citeauthoryear{{Wampler}, {Scargle} \& {Miller}}{{Wampler}
  et~al.}{1969}]{Wampler1969}
{Wampler} E.~J.,  {Scargle} J.~D.,    {Miller} J.~S.,  1969, \apjl, 157, L1

\bibitem[\protect\citeauthoryear{{Weisskopf}, {Silver}, {Kestenbaum}, {Long} \&
  {Novick}}{{Weisskopf} et~al.}{1978}]{Weisskopf1978}
{Weisskopf} M.~C.,  {Silver} E.~H.,  {Kestenbaum} H.~L.,  {Long} K.~S.,
  {Novick} R.,  1978, \apjl, 220, L117

\bibitem[\protect\citeauthoryear{{Woltjer}}{{Woltjer}}{1957}]{Woltjer1957}
{Woltjer} L.,  1957, \bain, 13, 301

\end{thebibliography}

\appendix
\onecolumn
\section{}

\label{Sec:Appendix} Here, we present step-by-step our
polarisation data analysis based on the  method described  by
\citet{Sparks1999}.  The  rotating polarisation filter (RPF)
inside the OPTIMA instrument  provides polarimetric data that
represent a series of `images' of an object taken through 180 sets
of linear polarisers, when we bin the continuous rotation of the
RPF into discrete one degree intervals. A single polariser is not
a 100\% perfect polariser, but its characteristics are well
established, and this is essential for the chosen data analysis
method. From an input data set of 180 idependent intensities
$I$ (measured in counts) and their errors ($\sigma_I = \sqrt I$),
corresponding to  a  set of observations through 180 identical but
not perfect polarisers we derive the Stokes parameters,
following the case of $n$ polarisers after \citet{Sparks1999}.

Linearly  polarised   light  requires  measurement  of   three  quantities
to  be fully characterised. There  are various ways  of expressing this. The
most  common  one involves  the total intensity of the light $I$, the degree of
polarisation $p$, and  the position angle $\theta$. An intermediate stage
between the input data  and  the solution  of polarisation quantities  are
the  Stokes parameters ($I$, $Q$, $U$) that are related through:
\begin{equation} \label{Eq:QU}
Q = I p \cos 2\theta, \qquad
U = I p \sin 2\theta,
\end{equation} or  equivalently,
\begin{equation} \label{Eq:pdpa}
p = \frac{(Q2+U2)^{1/2}}{I}, \qquad
\theta = \frac{1}{2} \arctan \left(\frac{U}{Q}\right).
\end{equation}
These quantities describe all intrinsic properties of linearly
polarised radiation. We would like to  underline that they
should not be confused with the properties  of the polarising
elements  of the polarimeter, which  in our case are the
properties of the measured intensities in each  one degree
intervals of the RPF, i.e. in each of 180 polarisers. These three
quantities that characterise fully the behaviour or response of
the linearly  polarising element are:
\begin{itemize}
\item[-] its overall throughput (hereafter $t$), in particular to unpolarised
light,
\item[-] its  efficiency as  a  polariser  (hereafter  $\epsilon$), i.e. the
ability to reject  and accept polarised light of  perpendicular and parallel
orientations,
\item[-] the position angle of the polariser (hereafter $\phi$).
\end{itemize}
There are a  variety of conventions commonly used  to present these quantities
\citep{Mazzuca1998}. Here,  we adopt the convention after \citealt{Sparks1999},
i.e. the output intensity of  a beam  with input  Stokes parameters  ($I$, $Q$,
$U$) passing  through a polarising element is given by
\begin{equation} \label{Eq:Ik}
I_k = \frac{1}{2} t_k [I + \epsilon_k (\cos 2\phi_k Q + \sin 2\phi_k U)]
\end{equation}
where $t_k$ is related to the throughput to unpolarised light,
$\epsilon_k$ is the  efficiency of  the polariser,  and $\phi_k$
is the position angle  of the polariser $k$. During our
measurements we always use the same Polaroid, just the position
angle of the filter changes, therefore $\epsilon_k = \epsilon =
0.998$, $t_k = t = 0.32$ (reference: Polarisation Filter Type VIS
4 K, Linos Photonics), and $\phi_k$ takes values from 0 to 179
with 1 degree steps.

Following the equations given by \citet{Sparks1999} we define the
new three-component vector  for the effective measurements:
\begin{align}
I''_1 = \sum\frac{I_k t_k}{{\sigma_k}^2}, \qquad I''_2 =
\sum\frac{I_k t_k \epsilon_k \cos 2 \phi_k}{{\sigma_k}^2}, \qquad
I''_3 = \sum\frac{I_k t_k \epsilon_k \sin 2 \phi_k}{{\sigma_k}^2}.
\end{align}

\begin{figure}
\centering
\includegraphics[scale=0.425]{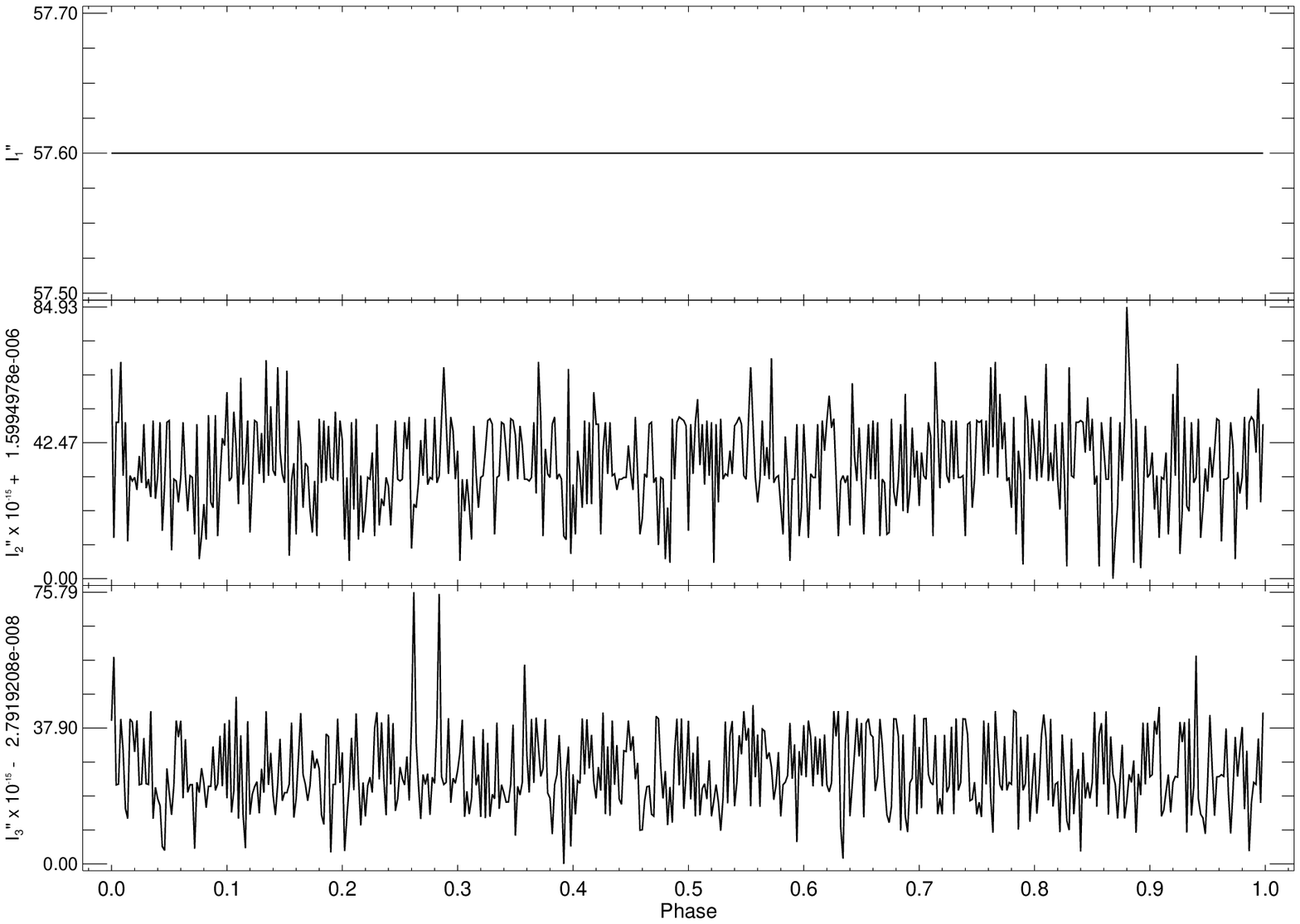}
\includegraphics[scale=0.425]{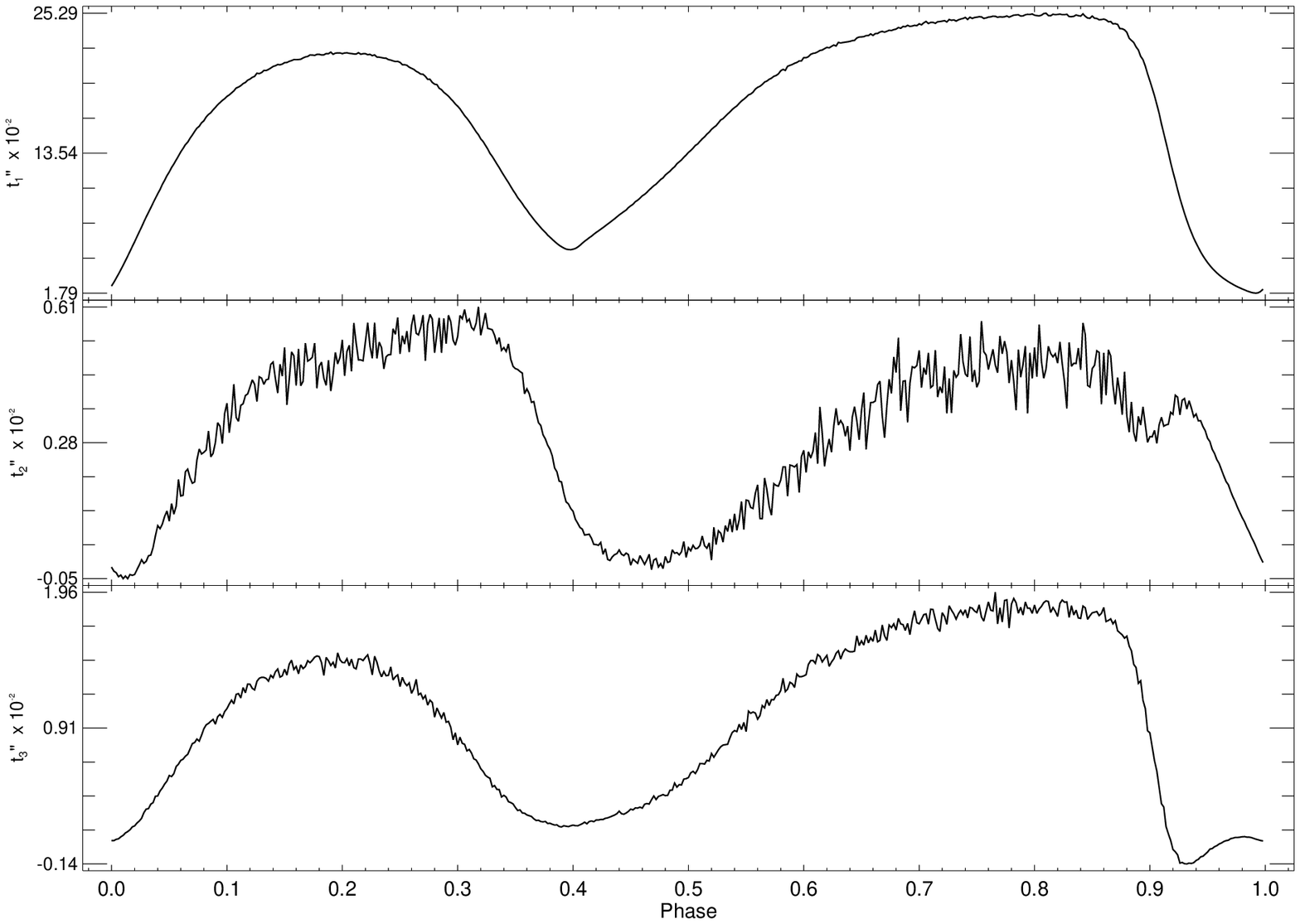}
\includegraphics[scale=0.425]{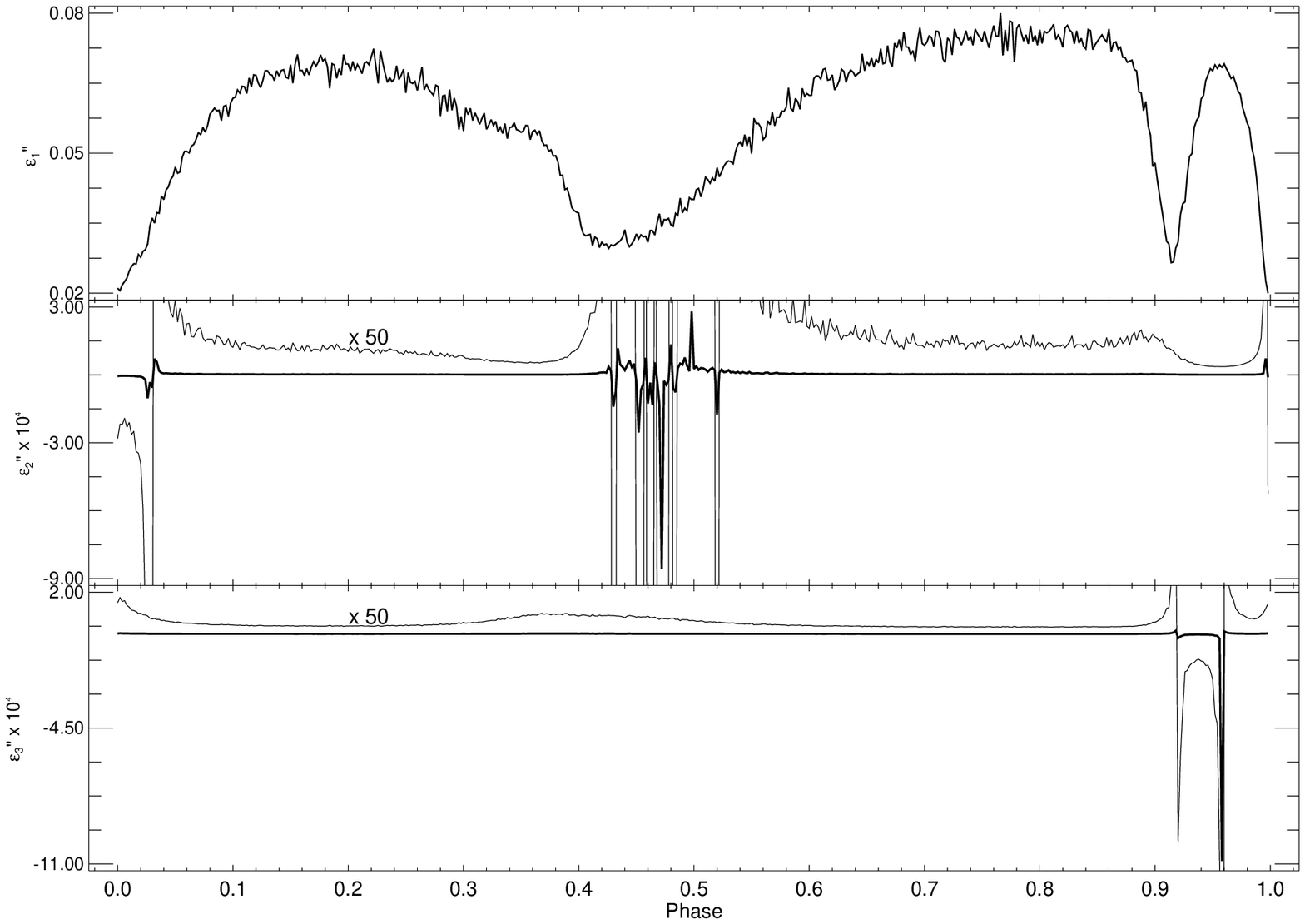}
\includegraphics[scale=0.425]{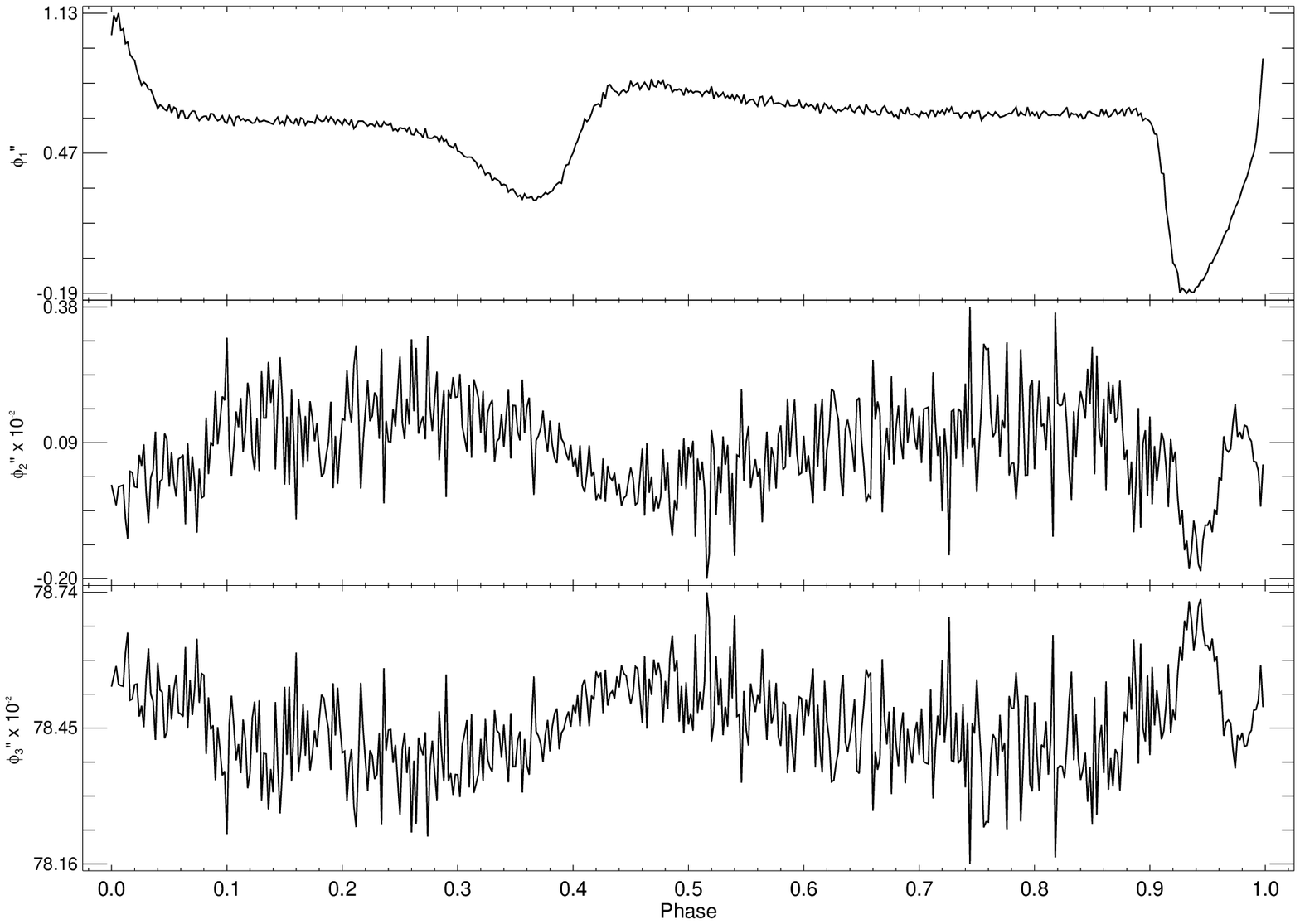}
\caption{From left to right: effective intensity  $I''_1, I''_2$,
$I''_3$, transmittance  $t''_1, t''_2$, $t''_3$,
efficiency  $\epsilon''_1, \epsilon''_2$, $\epsilon''_3$,
and position angle $\phi''_1, \phi''_2$, $\phi''_3$
derived from 180 polarisers as a function of the Crab
pulsar rotational phase (1 cycle = 500 bins).}
\label{Fig:i_prim}
\end{figure}

The error estimate for each measured intensity (=counts) is based
on Gaussian statistics; therefore $I_k={\sigma_k}^2$ and the
corresponding factors reduce to unity. The results  of calculating
the effective intensity  components $I''_1, I''_2$ and $I''_3$ are
shown in Fig.~\ref{Fig:i_prim}. In our  case these values are
constant. $I''_1$  is just a  sum over the  same 180 polarisers
with  the same throughput to  unpolarised light, i.e. $t=  0.32$.
Therefore we  get: $I''_1 = 180 \times 0.32 = 57.6$. Whereas,
$I''_2= 180 t \epsilon \sum{\cos 2 \phi_k}$ and $I''_3= 180 t
\epsilon \sum{\sin 2 \phi_k}$, where $\sum$ denotes a sum over
index  $k$ of the 180  polarisers. Knowing that $\epsilon = 0.998$
and the integrals of $\cos 2 \phi_k$ and $\sin  2 \phi_k$ over the
range from $0\degr$ to $180\degr$ we  expect  $I''_2 \sim  I''_1  \sum{\cos 2
\phi_k}$  and  $I''_3 \sim  I''_1 \sum{\sin 2 \phi_k}$,  within
available numerical precision, to  be very close to zero.

Similarly, we can define a three-component vector of effective transmittances
\begin{align}
t''_1 = \sum\frac{{t_k}^2}{{\sigma_k}^2}, \qquad t''_2 =
\sum\frac{{t_k}^2 \epsilon_k \cos 2 \phi_k}{{\sigma_k}^2}, \qquad
t''_3 = \sum\frac{{t_k}^2 \epsilon_k \sin 2 \phi_k}{{\sigma_k}^2},
\end{align}

and the vector of effective efficiencies
\begin{small}
\begin{align}
\epsilon''_1                          &=                         \frac{1}{\sum
  {t_k}^2/{\sigma_k}^2}
\sqrt{{\left(\sum\frac{{t_k}^2}{{\sigma_k}^2}   \epsilon_k  \cos
  2\phi_k  \right)}^2 +  {\left(  \sum\frac{{t_k}^2}{{\sigma_k}^2}\epsilon_k \sin  2
  \phi_k \right)}^2}, \nonumber\\
\epsilon''_2    &=   \frac{1}{\sum   {t_k}^2    \epsilon_k   \cos
2\phi_k   /
  {\sigma_k}^2}
\sqrt{{\left(\sum\frac{{t_k}^2}{{\sigma_k}^2}    {\epsilon_k}^2
\cos^2
  2\phi_k \right)}^2  + {\left( \sum\frac{{t_k}^2}{{\sigma_k}^2}{\epsilon_k}^2  \sin 2
  \phi_k \cos 2 \phi_k
\right)}^2}, \nonumber\\
\epsilon''_3    &=   \frac{1}{\sum   {t_k}^2    \epsilon_k   \sin
2\phi_k   /
  {\sigma_k}^2}
\sqrt{{\left(\sum\frac{{t_k}^2}{{\sigma_k}^2}     {\epsilon_k}^2
\sin
  2\phi_k         \cos        2\phi_k        \right)}^2         +        {\left(
  \sum\frac{{t_k}^2}{{\sigma_k}^2}{\epsilon_k}^2        \sin^2        2       \phi_k
  \right)}^2}, \nonumber
\end{align}
\end{small}

as well as a vector of effective position angles
\begin{align}
\phi''_1   &=  \frac{1}{2}   \arctan   \left(  \sum
\frac{{t_k}^2}{{\sigma_k}^2} \epsilon_k \sin  2\phi_k \Bigg/ \sum
\frac{{t_k}^2}{{\sigma_k}^2} \epsilon_k \cos
2\phi_k \right), \nonumber\\
\phi''_2        &=        \frac{1}{2}        \arctan        \left(
\sum \frac{{t_k}^2}{{\sigma_k}^2}{\epsilon_k}^2  \sin 2\phi_k
\cos 2\phi_k \Bigg/  \sum \frac{{t_k}^2}{{\sigma_k}^2}
{\epsilon_k}^2 \cos^2     2\phi_k     \right),
\nonumber\\
\phi''_3   &=  \frac{1}{2}   \arctan   \left(  \sum
\frac{{t_k}^2}{{\sigma_k}^2} {\epsilon_k}^2 \sin^2 2\phi_k  \Bigg/
\sum \frac{{t_k}^2}{{\sigma_k}^2} {\epsilon_k}^2 \sin 2\phi_k \cos
2\phi_k \right). \nonumber
\end{align}

\begin{figure}
\centering
\includegraphics[scale=0.6]{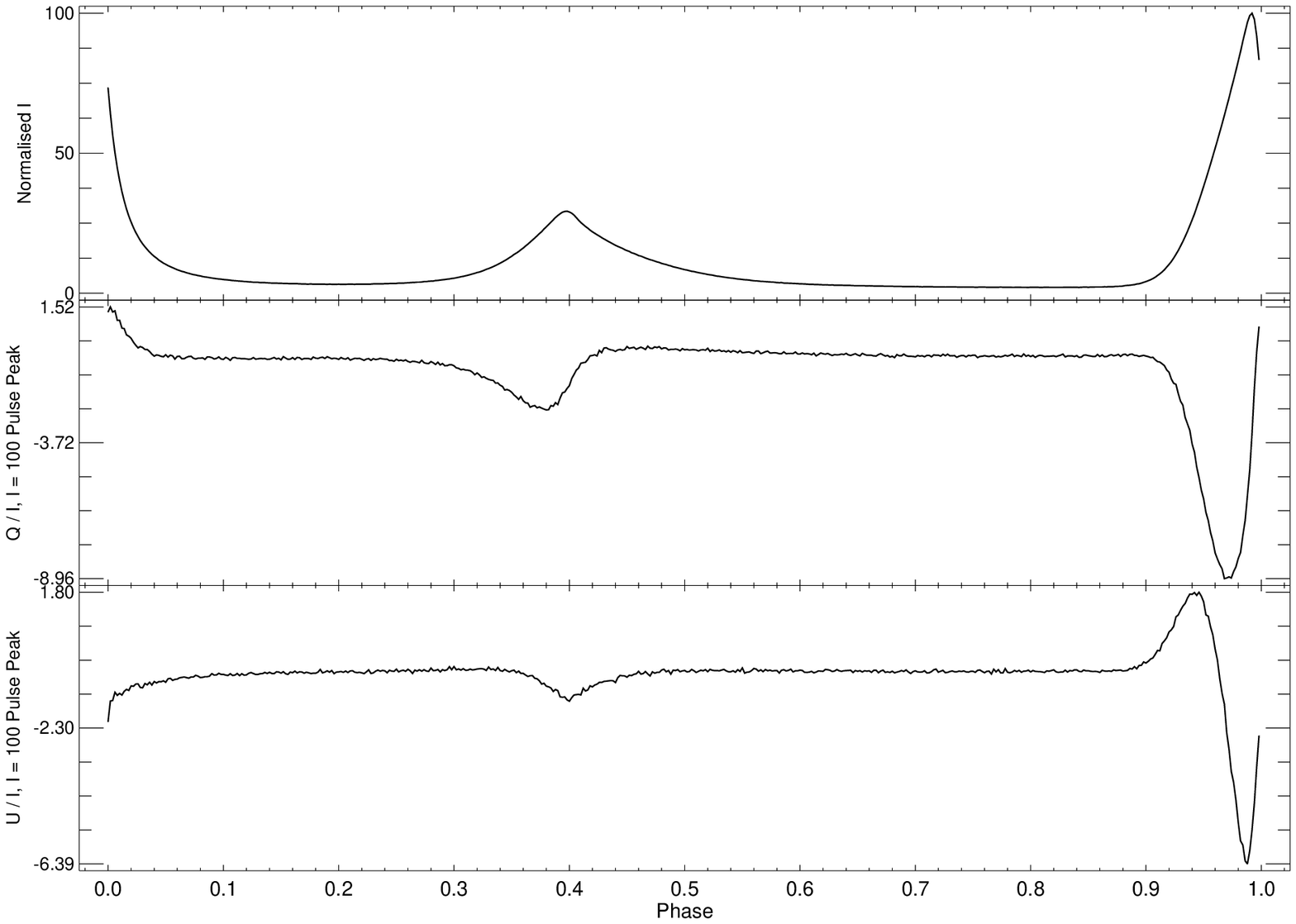}
\caption{Stokes parameters $I$, $Q$,  $U$ derived from
180 polarisers as a function   of   the  Crab   pulsar   rotation
phase   (1  cycle   =   500 bins). Normalisation was  such that
$I=100$ at the maximum,  and $Q$ and $U$ are normalised to  $I$.
These values corresponds to the  values shown on the QU plane in
the Fig.~\ref{Fig:pa_pd}.}
\label{Fig:stokes_norm}
\end{figure}

The effective intensity, transmittances, efficiencies, and
position angles are shown in  Fig.~\ref{Fig:i_prim}. By making
these substitutions, the solution for the Stokes vector is given
by
\begin{equation}
(I, Q, U) = B
  \left(\begin{array}{c}I''_{1}/(0.5t''_1)\\I''_{2}/(0.5t''_2)\\I''_{3}/(0.5t''_3)
\end{array}\right)
  \end{equation}
where
\begin{equation} \label{Eq:b}
B = \left(\begin{array}{ccc}
\epsilon''_2 \epsilon''_3 \sin(2\phi''_3 - 2\phi''_2) &
\epsilon''_1 \epsilon''_3 \sin(2\phi''_1 - 2\phi''_3) &
\epsilon''_1 \epsilon''_2 \sin(2\phi''_2 - 2\phi''_1) \\
\epsilon''_2 \sin 2\phi''_2 - \epsilon''_3 \sin 2\phi''_3 &
\epsilon''_3 \sin 2\phi''_3 - \epsilon''_1 \sin 2\phi''_1 &
\epsilon''_1 \sin 2\phi''_1 - \epsilon''_2 \sin 2\phi''_2 \\
\epsilon''_3 \cos 2\phi''_3 - \epsilon''_2 \cos 2\phi''_2 &
\epsilon''_1 \cos 2\phi''_1 - \epsilon''_3 \cos 2\phi''_3 &
\epsilon''_2 \cos 2\phi''_2 - \epsilon''_1 \cos 2\phi''_1
\end{array}\right) / \Omega
\end{equation}
and
\begin{equation}
\Omega = \epsilon''_1 \epsilon''_2 \sin(2\phi''_2 - 2\phi''_1) +
\epsilon''_2 \epsilon''_3 \sin(2\phi''_3 - 2\phi''_2) +
\epsilon''_1 \epsilon''_3 \sin(2\phi''_1 - 2\phi''_3).
\end{equation}
can be  used immediately. In this way  we obtained the Stokes
parameters shown in Fig.~\ref{Fig:stokes_norm}.

\begin{figure*}
\centering
\includegraphics[scale=0.94]{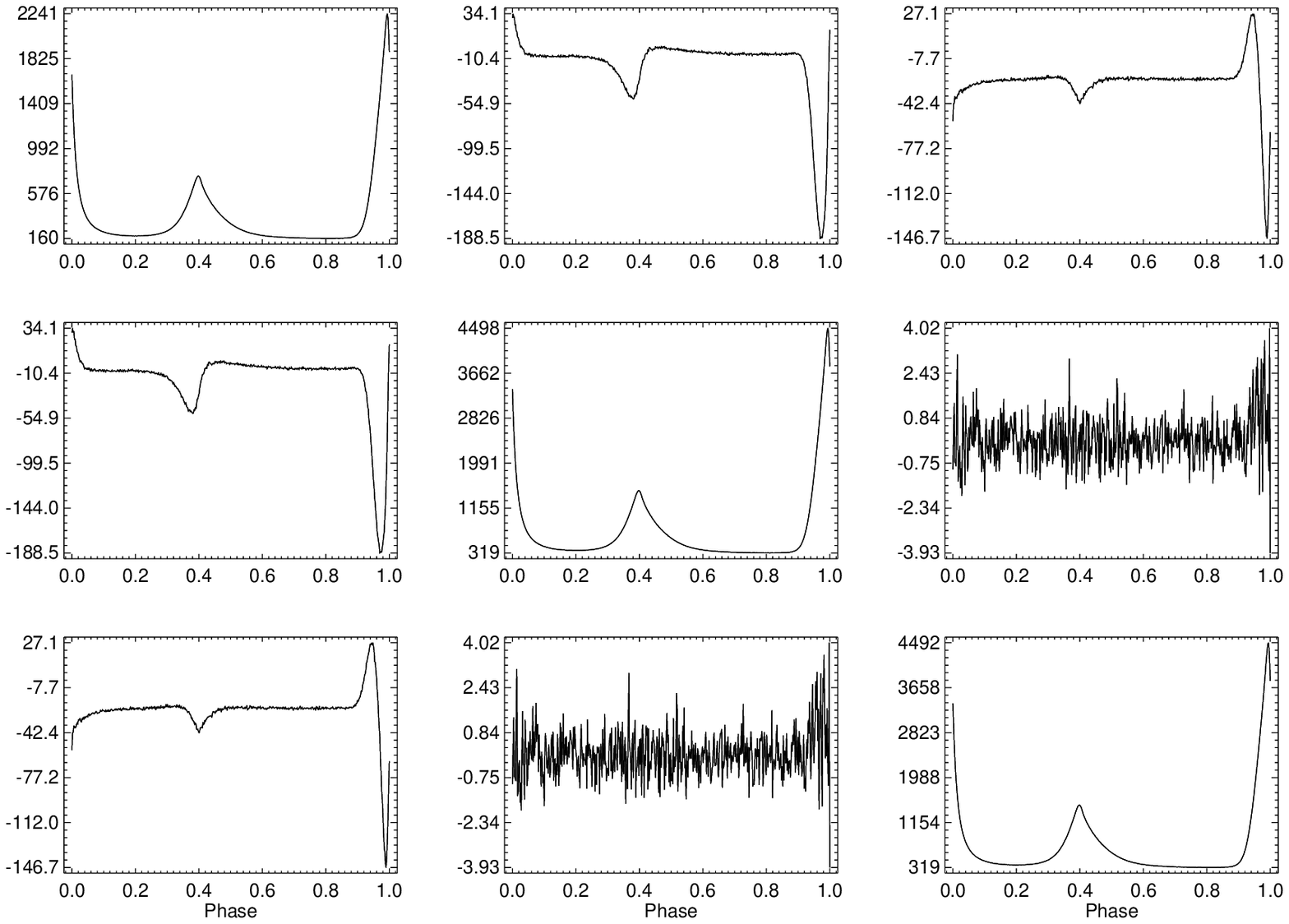}
\caption{Covariance matrix as a function of the Crab pulsar rotational phase
(1 cycle = 500 bins).}
\label{Fig:cov_matrix}
\end{figure*}

We calculated the  covariance matrix (Fig.~\ref{Fig:cov_matrix}) defined
as the inverse of matrix $C$ given by Eq.~8 in \citet{Sparks1999}. Following
the  error  propagation  equation   we  calculate  the  uncertainties  of  the
polarisation degree and the position angle according to:

\begin{align}
\sigma_{p}^2 &\simeq (\frac{\partial p}{\partial Q})^2 \sigma_{Q}^2 + 
(\frac{\partial p}{\partial U})^2 \sigma_{U}^2 + 
(\frac{\partial p}{\partial I})^2 \sigma_{I}^2 
+ 2\sigma_{QI}^2(\frac{\partial p}{\partial Q}) (\frac{\partial p}{\partial I}) +
2\sigma_{UI}^2(\frac{\partial p}{\partial U}) (\frac{\partial p}{\partial I})
+ 2\sigma_{QU}^2(\frac{\partial p}{\partial Q}) (\frac{\partial p}{\partial U})
\end{align}

\begin{equation}
\sigma_{\theta}^2 \simeq \sigma_{U}^2 (\frac{\partial \theta}{\partial U})^2 + 
\sigma_{Q}^2 (\frac{\partial \theta}{\partial Q})^2 +
2\sigma_{UQ}^2 (\frac{\partial \theta}{\partial U})(\frac{\partial \theta}{\partial Q}).
\end{equation}

where appropriate standard deviations are the components of the covariance  matrix.
In general the diagonal terms dominate the uncertainties. Covariant terms make the
maximum contribution on the level of 5\% and 1\% to the $\sigma_{p}$ and
$\sigma_{\theta}$, respectively.

\label{lastpage}

\end{document}